# Forecasting and Explaining the Phillips Curve: A SHAP-Based Comparison of Machine Learning and Traditional Time-Series Models for Canadian Unemployment and Inflation

**Louis Agyekum[1*]**

[1*]Department of Economics, University of Ottawa, Ottawa, ON, K1N 6N5, Canada

***Correspondence: Email: agyekumlouis9@gmail.com**

## Abstract

This study evaluates the out-of-sample forecasting ability of six model types: ARIMA, VAR, Random Forest, XGBoost, LSTM, and GRU, for monthly Canadian inflation from January 2012 to April 2026 (n = 172). The evaluation employs expanding-window walk-forward validation across 1-, 3-, 6-, and 12-month horizons. Results reveal a horizon-dependent shift: ARIMA significantly outperforms all machine learning and deep learning models at the one-month horizon (Diebold-Mariano $p < 0.05$). However, Random Forest and XGBoost become notably superior at six and twelve months, reducing RMSE by approximately 30–75% compared to ARIMA and VAR. LSTM and GRU perform well only at the shortest horizon, likely due to overfitting given the limited data. Analyzing four macroeconomic sub-periods shows that no single model consistently dominates. SHAP analysis of the top-performing XGBoost model indicates that lagged inflation is more influential than unemployment, which only becomes significantly impactful during the pandemic tail. The findings clarify when machine learning methods can surpass traditional benchmarks.

## 1. Introduction

The relationship between unemployment and inflation, known as the Phillips curve, remains one of the most empirically contested and policy-relevant in macroeconomics. Since the 2008 financial crisis, a substantial literature has debated whether this relationship has flattened, become nonlinear, or simply become harder to estimate using conventional linear specifications (Hooper et al., 2020; Hazell et al., 2022). At the same time, the rapid adoption of machine learning (ML) and deep learning (DL) methods in empirical macroeconomics has produced a parallel literature asking whether flexible, nonlinear, data-driven models can outperform traditional time-series benchmarks such as the autoregressive integrated moving average (ARIMA) model and the vector autoregression (VAR) in forecasting key macroeconomic aggregates (Medeiros et al., 2021; Coulombe et al., 2022; Nortey et al., 2025).

These two literatures have largely developed in parallel. Forecasting comparisons between ML/DL models and traditional benchmarks typically report a single point estimate of relative accuracy, a horse race with a declared winner, without assessing whether that ranking is stable across forecast horizons or macroeconomic regimes, nor linking the model's internal structure to the economic relationship of interest. Conversely, the Phillips curve flattening literature has relied primarily on linear or piecewise-linear specifications, which, by construction, cannot reveal whether the true unemployment–inflation relationship is smoothly nonlinear, threshold-based, or asymmetric across the business cycle.

This paper bridges these two strands of research, drawing on 42 years of theoretical antecedents to motivate three connected empirical exercises. First, we conduct a rigorous, horizon- and regime-conditional forecasting comparison of six model classes (ARIMA, VAR, Random Forest, XGBoost, LSTM, and GRU) for Canadian inflation, evaluated using an expanding-window walk-forward design that mimics how a real-time forecaster would operate. Second, rather than reporting a single global ranking, we ask whether forecasting accuracy is conditional on the forecast horizon (1, 3, 6, and 12 months ahead) and on the prevailing macroeconomic regime (the pre-pandemic period, the 2020 COVID-19 shock, the 2021–2023 inflation surge, and the subsequent disinflation). Third, and most centrally, we apply Shapley Additive exPlanations (SHAP; Lundberg & Lee, 2017) to the best-performing tree-based model to recover an implied, nonlinear, data-driven Phillips curve, and we ask how the shape of this implied relationship compares to the piecewise-linear flattening narrative that dominates the existing macroeconomic literature.

Canada offers a useful, if modest-sized, laboratory for this exercise. The sample period (January 2012 to April 2026, n = 172 monthly observations) spans a long pre-pandemic period of low and stable inflation, an abrupt pandemic-era demand and supply shock, a sharp and sustained inflation surge that pushed year-over-year Consumer Price Index (CPI) inflation above 8% in 2022, and a subsequent disinflation back toward the Bank of Canada's 2% target. This sequence of distinct regimes, compressed into a relatively short sample, provides a natural test bed for assessing whether model rankings are regime-dependent, a question that is difficult to answer convincingly with longer but more homogeneous historical samples.

Our central forecasting finding is a clear horizon-dependent crossover. At the one-month-ahead horizon, ARIMA is statistically more accurate than every machine learning and deep learning model considered, a result confirmed by Diebold-Mariano tests at conventional significance levels. At longer horizons of six and twelve months, however, this ranking reverses: Random Forest and XGBoost produce substantially lower forecast errors than both ARIMA and VAR, while the recurrent neural network models (LSTM, GRU) are competitive only in the short run and deteriorate markedly at longer horizons. We interpret this pattern as evidence that the relative advantage of flexible, nonlinear models in macroeconomic forecasting is not uniform across horizons. Very short-horizon inflation dynamics in this sample remain well captured by a parsimonious linear autoregressive structure, whereas longer-horizon dynamics benefit from the ability of tree-based ensembles to capture nonlinear interactions among lagged unemployment and inflation.

Our SHAP-based interpretability results offer a structural lens on this forecasting exercise. Own-lags of inflation dominate the model's feature importance ranking, consistent with strong inflation persistence and inertia, while lagged unemployment has smaller but economically meaningful importance. The SHAP-implied relationship between lagged unemployment and predicted inflation is negative and relatively flat across most of the historically observed unemployment range (approximately 5% to 8%), but steepens markedly in the extreme upper tail of the unemployment distribution associated with the 2020 pandemic shock (above approximately 10%). This finding materially differs from the canonical "flattening" narrative, which centers on changes in the slope at low, tight-labor-market unemployment rates (Hooper et al., 2020); our results instead point to nonlinearity concentrated in extreme, infrequent, high-unemployment episodes, a pattern that is robust to the choice of tree-based estimator and to reversing the direction of prediction.

The remainder of the paper is organized as follows. Section 2 reviews the literature on Phillips curve estimation and machine learning for macroeconomic forecasting. Section 3 describes the data. Section 4 details the econometric and machine learning methodology, including the walk-forward validation design, the Diebold-Mariano test, and the SHAP framework. Section 5 presents the empirical results. Section 6 discusses the economic interpretation of the findings. Section 7 acknowledges limitations, and Section 8 concludes.

## 2. Literature Review

### 2.1 The Phillips Curve and Its Stability

The empirical Phillips curve, the negative relationship between unemployment and wage or price inflation, has been a central organizing concept in monetary economics since its introduction, yet its stability and estimability have repeatedly been questioned. Hooper et al. (2020) provide a comprehensive review of the evidence on whether the Phillips curve has "flattened" in recent decades, concluding that national-level United States data since the 1980s show substantially weaker evidence of both a negative slope and nonlinearity than in earlier decades, while state- and metropolitan-level data continue to show significant negative slopes and steeper relationships, particularly in tight labor markets (that is, at low unemployment rates). This finding that nonlinearity in the Phillips curve, where it exists, tends to manifest as a steeper slope at low unemployment rather than at high unemployment is an important point of comparison for the SHAP-based results presented in this paper.

Related work by Hazell et al. (2022) uses U.S. state-level data to argue that the slope of the price Phillips curve has historically been small, even before the 2010s, suggesting that apparent flattening at the national level partly reflects identification challenges stemming from the endogeneity of monetary policy rather than a genuine change in the underlying structural relationship. Smith, Timmermann, and Wright (2023) examine breaks in the Phillips curve using panel data methods and similarly find a flattening break in the price Phillips curve, with a smaller break in the wage Phillips curve. The Federal Reserve Board's own research has further emphasized the possibility of a kinked or threshold-based nonlinearity, with several studies finding the curve to be steeper once unemployment falls below a threshold of roughly 4 to 5 percent.

A common feature of this literature is its reliance on linear, piecewise-linear, or low-order polynomial specifications estimated by ordinary least squares or related parametric methods. While computationally and conceptually convenient, these approaches impose a functional form on the unemployment–inflation relationship before estimation, which may obscure genuinely data-driven nonlinearities, particularly in the tail regions of the unemployment distribution, which are sparsely populated in any single country's time series. This motivates the use of nonparametric, tree-based machine learning models combined with post hoc interpretability tools, as pursued in this paper.

## 2.2 Machine Learning and Deep Learning in Macroeconomic Forecasting

A growing literature evaluates whether machine learning methods can improve upon traditional linear time-series models for forecasting macroeconomic aggregates. Medeiros et al. (2021) compare a wide range of ML methods against standard linear benchmarks for forecasting U.S. inflation and find that Random Forest-based models deliver substantial accuracy gains over the most common univariate and multivariate linear models, particularly at intermediate forecast horizons. However, the relative ranking of models is sensitive to the specific inflation measure and time period considered. Coulombe et al. (2022) provide a broader assessment of when machine learning is and is not useful for macroeconomic forecasting, arguing that the benefits of ML methods are concentrated in settings with genuine nonlinearity or regime-switching behavior, and that gains are smaller or absent for series that are well approximated by linear models over the sample period considered.

Most directly relevant to this paper is Nortey et al. (2025), who compare deep learning models (including transformer architectures), machine learning models (gradient boosting, XGBoost,

AdaBoost, and linear regression), and traditional statistical benchmarks for forecasting monthly U.S. inflation. Their study, like the present one, places deep learning and machine learning forecasts directly against classical time-series benchmarks rather than against each other in isolation, and frames the central question as whether the additional flexibility of deep learning architectures translates into superior forecasting accuracy for inflation specifically, rather than for macroeconomic series more generally. The present paper extends this line of inquiry in three ways: by evaluating forecast accuracy conditional on forecast horizon and macroeconomic regime rather than reporting a single aggregate ranking; by applying the comparison to Canadian rather than U.S. data, in a sample that spans a particularly turbulent set of regimes (the COVID-19 shock and the subsequent global inflation surge); and by pairing the forecasting comparison with a SHAP-based structural interpretation of the winning model, explicitly connecting the forecasting and interpretability literatures.

Beyond Nortey et al. (2025), several related contributions document mixed evidence on the relative performance of machine learning and deep learning models in macroeconomic forecasting applications. Several papers report that simple tree-based ensembles are difficult to beat when comparisons are made conditional on horizon, when a rigorous out-of-sample validation design is used, or when model complexity is fairly accounted for relative to sample size. This concern is especially salient for the deep learning architectures evaluated in this paper, given the relatively short (n = 172) Canadian sample.

### 2.3 Explainable Machine Learning and SHAP

The growing use of complex, nonlinear models in economics and finance has been accompanied by rising demand for post hoc interpretability methods that help researchers understand what a trained model has learned. Lundberg and Lee (2017) introduce Shapley Additive explanations (SHAP), a unified framework grounded in cooperative game theory that assigns each input feature an additive contribution to a given prediction and satisfies a set of desirable theoretical properties, including local accuracy, consistency, and missingness. For tree-based ensemble models such as Random Forest and XGBoost, SHAP values can be computed exactly and efficiently via the Tree Explainer algorithm, which exploits the structure of decision trees to avoid the exponential computational cost of the general Shapley value calculation.

In macroeconomic applications, SHAP and related interpretability tools have begun to be used to recover implied structural relationships from otherwise opaque machine learning models. This approach treats the trained model as a flexible nonparametric estimator and uses SHAP to visualize the marginal relationship between a given predictor and the outcome of interest, conditional on all other predictors in the model. This paper applies this logic to the unemployment–inflation relationship, treating the SHAP dependence structure between lagged unemployment and predicted inflation as an implied, data-driven analogue to the structural Phillips curve, and comparing its shape to the piecewise-linear flattening evidence reviewed in Section 2.1.

### 2.4 Contribution

This paper contributes to the literature in three specific ways. First, it provides a horizon- and regime-conditional comparison of six model classes for forecasting Canadian inflation, addressing the methodological concern that single aggregate accuracy rankings can mask substantial heterogeneity in relative performance across forecasting horizons and macroeconomic conditions. Second, it applies formal statistical tests of equal predictive accuracy (Diebold & Mariano, 1995) to every pairwise model comparison, rather than relying solely on point estimates of forecast error.

Third, and most distinctively, it uses SHAP to recover an implied nonlinear Phillips curve directly from the best-performing machine learning model and situates the shape of this implied relationship explicitly within the existing literature on flattening and nonlinearity, providing a bridge between the macro-forecasting and structural-interpretation literatures that have so far developed largely independently of one another.

3. Data and Methodology

The empirical analysis uses two monthly Canadian macroeconomic time series covering January 2012 to April 2026 (172 observations): the seasonally unadjusted unemployment rate for the total population aged 15 and over, sourced from Statistics Canada's Labour Force Survey (Table 14-10-0374-01), and the year-over-year percentage change in the all-items Consumer Price Index, sourced from Statistics Canada's Consumer Price Index tables (Table 18-10-0004-01, 2002 = 100). The inflation series is computed as the twelve-month percentage change in the price index, the standard definition used by Statistics Canada and the Bank of Canada in official communications. Both series are merged onto a common monthly date index, with no missing observations across the full sample window after the initial 12 months are dropped to compute the year-over-year inflation rate.

**4.1 Stationarity and Diagnostic Testing**

Before model estimation, both series are tested for stationarity using three complementary unit root tests with distinct null hypotheses to triangulate the results. The Augmented Dickey-Fuller (ADF) test estimates the regression

$$\Delta y_t = \alpha + \beta y_{t-1} + \sum_{i=1}^{p} \gamma_i \, \Delta y_{t-i} + \varepsilon_t \qquad (1)$$

where the null hypothesis $H_0$: $\beta = 0$ indicates the presence of a unit root (non-stationarity), with the lag order $p$ selected via the Akaike Information Criterion (AIC). The Kwiatkowski-Phillips-Schmidt-Shin (KPSS) test reverses this null, testing $H_0$: the series is stationary against the alternative of a unit root, using the test statistic

$$LM = \frac{1}{T^2} \sum_{t=1}^{T} \frac{S_t^2}{\hat{\sigma}^2} \qquad (2)$$

where $S_t$ is the partial sum of residuals from a regression of the series on a constant (and, where specified, a trend), and $\hat{\sigma}^2$ is a consistent estimator of the long-run variance. The Phillips-Perron (PP) test provides a third, nonparametric correction for serial correlation and heteroskedasticity in the error term of equation (1), while retaining the same null hypothesis as the ADF test. Because the ADF and PP tests share a null of non-stationarity, whereas KPSS reverses it, results across the three tests are interpreted jointly: a series is treated as more confidently stationary when ADF and PP reject their null and KPSS fails to reject its null.

Residual autocorrelation is assessed via the Ljung-Box Q-statistic,

$$Q = T(T+2) \sum_{k=1}^{h} \frac{\hat{\rho}_k^2}{T-k} \qquad (3)$$

Tests are conducted up to lag $h = 12$, with the null hypothesis of no autocorrelation up to the specified lag. Conditional heteroskedasticity is assessed using the ARCH-LM test of Engle (1982), which regresses squared residuals on their lags and tests the joint significance of the resulting coefficients using an LM statistic that is asymptotically $\chi^2$ under the null of no ARCH effects.

### 4.2 Causality and Cointegration

Granger causality between unemployment and inflation is assessed by estimating, for each direction and for lag orders from one to twelve, the restricted and unrestricted regressions

$$y_t = \alpha + \sum_{i=1}^{p} \beta_i \, y_{t-i} + \sum_{i=1}^{p} \gamma_i \, x_{t-i} + \varepsilon_t \qquad (4)$$

and testing the joint null hypothesis $H_0$: $\gamma_1 = \gamma_2 = \ldots = \gamma_p = 0$ via an F-test, where rejection indicates that lagged values of *x* improve the prediction of *y* beyond what is achieved using lagged values of *y* alone (Granger causality in the statistical, not necessarily structural, sense). Long-run equilibrium between the two series is assessed using the Johansen (1991) trace test for cointegration, which tests the rank *r* of the cointegrating relationship in a vector error-correction representation of the bivariate system, with critical values from the asymptotic distribution tabulated by Johansen and Juselius (1990).

### 4.3 Structural Break Detection

Structural breaks in each series are identified using two complementary approaches. First, the Pruned Exact Linear Time (PELT) algorithm of Killick, Fearnhead, and Eckley (2012) is applied to each series to minimize a penalized sum-of-costs criterion.

$$\min \sum_{i=1}^{m+1} \left[ C\left( y_{t_{i-1}+1:T_i} \right) \right] + \beta \cdot m \qquad (5)$$

over the number and location of breakpoints *m*, where C(·) is a segment-specific cost function (here, based on a radial basis function kernel to allow for nonlinear within-segment variation) and β is a penalty term that controls the number of detected breaks. The PELT algorithm achieves the global minimum of this criterion with a computational cost that is linear in the sample size under mild conditions, an improvement over the dynamic-programming approach of Bai and Perron (2003), which we use as a conceptual reference point for the underlying multiple-structural-change framework. Second, as a complementary check, the cumulative sum (CUSUM) test is applied to the recursive residuals of a simple first-order autoregressive benchmark for each series, testing the null hypothesis of parameter stability over the full sample against the alternative of one or more unmodeled structural breaks.

### 4.4 Traditional Time-Series Benchmarks

Two traditional time-series models serve as benchmarks for forecasting. The univariate ARIMA(p,d,q) model for inflation $\pi_t$ is specified as

$$\Phi(L)(1-L)^d\pi_t = \Theta(L)\varepsilon_t \qquad (6)$$

where $\Phi(L)$ and $\Theta(L)$ are lag polynomials of orders *p* and *q*, respectively; *L* is the lag operator; and *d* is the order of differencing required to achieve stationarity. The orders (p, d, q) are selected via a grid search over $p, q \in \{0, \ldots, 4\}$ and $d \in \{0, 1\}$, choosing the combination that minimizes the Akaike Information Criterion. The bivariate vector autoregression of order *k*, VAR(k), is specified as

$$Y_t = c + A_1Y_{t-1} + A_2Y_{t-2} + \cdots + A_kY_{t-k} + \varepsilon_t \qquad (7)$$

where $Y_t = [\text{unemployment}_t, \text{inflation}_t]'$ is the vector of the two endogenous variables, *c* is a vector of constants, $A_1, \ldots, A_k$ are coefficient matrices, and $\varepsilon_t$ is a white-noise error term. The lag order *k* is selected using the Akaike Information Criterion over candidate orders from one to twelve.

### 4.5 Machine Learning and Deep Learning Models

Two tree-based ensemble methods, Random Forest and Extreme Gradient Boosting (XGBoost), and two recurrent neural network architectures, long short-term memory (LSTM) and gated recurrent unit (GRU) networks, are evaluated alongside traditional benchmarks. For the tree-based models, the feature set consists of twelve lags (one through twelve months) of both the unemployment and inflation rates, together with calendar-month indicator variables to capture residual seasonality, for a total of up to thirty-five predictors after one-hot encoding. Random Forest aggregates predictions from an ensemble of regression trees grown on bootstrap-resampled subsets of the training data, with random feature subsampling at each split. It is specified here with 300 trees, a maximum depth of 6, and a minimum of 3 observations per leaf. XGBoost builds an additive ensemble of regression trees sequentially, with each new tree fit to the negative gradient of the loss function with respect to the current ensemble's predictions. It is specified here with 300 boosting rounds, a maximum tree depth of four, a learning rate of 0.03, and row and column subsampling rates of 0.8 to mitigate overfitting given the modest sample size.

The LSTM and GRU models use a sliding window of the preceding 12 months for both series (min-max scaled to the unit interval) as input. The input passes through a single recurrent layer with 32 units, a dropout layer (rate 0.2) for regularization, and a dense output layer. Given the modest sample size ($n \approx 172$), deliberately small and regularized architectures are used, together with early stopping based on training loss, to mitigate the well-documented risk of overfitting when deep learning models are applied to short macroeconomic time series.

**4.6 Walk-Forward Validation and Forecast Evaluation**

All six models are evaluated using an expanding-window walk-forward validation design, in which each model is re-estimated using only data available up to time $t$ and then used to generate forecasts for $t + h$, for horizons $h \in \{1, 3, 6, 12\}$ months. This design avoids look-ahead bias and more closely mimics the information set available to a real-time forecaster than a single train/test split. The initial training window comprises the first 60% of the sample (through August 2020); the remaining out-of-sample evaluation period (September 2020 through April 2026) spans the COVID-19 shock, the inflation surge, and the disinflation regimes, yielding 57 out-of-sample evaluation points at the one-, three-, and six-month horizons and 45 points at the twelve-month horizon, after accounting for the additional data required to construct longer-horizon targets. Tree-based models are re-estimated at every step of the expanding window; the recurrent neural network models are re-estimated every three months within the expanding window as a practical compromise given the computational cost of repeated neural network training, a common practice in applied deep learning forecasting studies.

Forecast accuracy is assessed using the root mean squared error (RMSE), mean absolute error (MAE), and mean absolute percentage error (MAPE):

$$RMSE = \sqrt{\frac{1}{n}\sum_{t=1}^{n}(y_t - \hat{y}_t)^2} \qquad (8)$$

$$MAE = \frac{1}{n}\sum_{t=1}^{n}|y_t - \hat{y}_t| \qquad (9)$$

$$MAPE = \frac{100}{n}\sum_{t=1}^{n}\left|\frac{y_t - \hat{y}_t}{y_t}\right| \qquad (10)$$

computed over the intersection of evaluation dates available across all six models at each horizon, ensuring a like-for-like comparison.

### 4.7 Diebold-Mariano Test

Lower point-estimate RMSE alone does not establish that one model is statistically more accurate than another. We therefore apply the Diebold and Mariano (1995) test for equal predictive accuracy to each pairwise comparison between the best-performing traditional benchmark (selected at each horizon as the one with the lower RMSE between ARIMA and VAR) and each of the four machine learning and deep learning challengers. Define the loss differential

$$d_t = \left(y_t - \hat{y}_{t,1}\right)^2 - \left(y_t - \hat{y}_{t,2}\right)^2 \qquad (11)$$

For forecasts from models 1 and 2 under squared-error loss. The Diebold-Mariano statistic is

$$DM = \frac{\bar{d}}{\sqrt{[\hat{\gamma}_0 + 2\sum_{k=1}^{h-1}\hat{\gamma}_k]/n}} \qquad (12)$$

where $\bar{d}$ is the sample mean of the loss differential, $\hat{\gamma}_k$ is the sample autocovariance of $d_t$ at lag *k*, and the summation accounts for serial correlation in the loss differential arising from overlapping multi-step-ahead forecasts at horizons greater than one. We apply the small-sample correction of Harvey, Leybourne, and Newbold (1997), which scales the test statistic by the factor $\sqrt{[(n + 1 - 2h + h(h-1)/n)/n]}$ before comparing it to a Student's *t* distribution with $n - 1$ degrees of freedom, a standard adjustment given the relatively small number of out-of-sample evaluation points, particularly at the twelve-month horizon.

### 4.8 SHAP Interpretability Analysis

To recover the economic relationship implied by the best-performing machine learning model, we apply Shapley Additive explanations (SHAP) to the final XGBoost model fit to the full sample at the one-month horizon. For a prediction f(x) generated from a feature vector x = ($x_1$, …, $x_m$), the SHAP value $\varphi_i$ for feature *i* is defined using the Shapley value from cooperative game theory,

$$\varphi_i = \sum_{S \subseteq M \setminus \{i\}} \left[\frac{|S|!\,(|M| - |S| - 1)!}{|M|!}\right] [f(S \cup \{i\}) - f(S)] \qquad (13)$$

where *M* is the full set of features, *S* ranges over all subsets of features excluding *i*, and f(S) denotes the model's expected prediction conditional on only the features in subset *S* being known. This formulation guarantees that the SHAP values satisfy local accuracy (the sum of feature contributions plus a baseline value equals the model's actual prediction for a given observation), consistency, and the missingness property, as established by Lundberg and Lee (2017). For tree-based ensembles, exact SHAP values are computed efficiently using the TreeExplainer algorithm, which exploits the recursive structure of decision trees to avoid the combinatorial cost of evaluating equation (13) directly. We use the SHAP dependence relationship between the one-month lag of the unemployment rate and its corresponding SHAP value, smoothed via locally weighted scatterplot smoothing (LOWESS), as the data-driven analogue to the structural Phillips curve. We compare its shape across the four macroeconomic regimes defined in Section 3 and against the existing flattening and nonlinearity literature reviewed in Section 2

## 5. Results

Table 1 reports descriptive statistics for both series over the full sample. The unemployment rate averages 6.77%, with a standard deviation of 1.42 percentage points, ranging from a minimum of 4.50% to a maximum of 14.30% (reflecting the acute pandemic-era spike in 2020). The inflation rate averages 2.29%, close to the Bank of Canada's 2% inflation target, with a standard deviation of 1.61 percentage points, ranging from −0.37% (a brief deflationary episode in 2020) to 8.13% (the peak of the 2022 inflation surge). Both series exhibit substantial positive skewness and excess kurtosis, and the Jarque-Bera test rejects normality for both series at any conventional significance level. This finding is unsurprising given the extreme and infrequent nature of the pandemic-era observations and motivates, in part, the use of nonparametric machine learning methods that do not impose distributional assumptions on the underlying data.

**Table 1.** Descriptive Statistics, January 2012–April 2026 (n = 172)

| Variable | Mean | Std. Dev. | Min | Max | Skewness | Jarque-Bera (p) |
|---|---|---|---|---|---|---|
| Unemployment rate (%) | 6.769 | 1.417 | 4.50 | 14.30 | 2.198 | 659.63 (<0.001) |
| Inflation rate (% YoY) | 2.287 | 1.610 | −0.37 | 8.13 | 1.692 | 137.62 (<0.001) |

*Note: Inflation rate is the year-over-year percentage change in the all-items Consumer Price Index. Source: Statistics Canada Tables 14-10-0374-01 and 18-10-0004-01.*

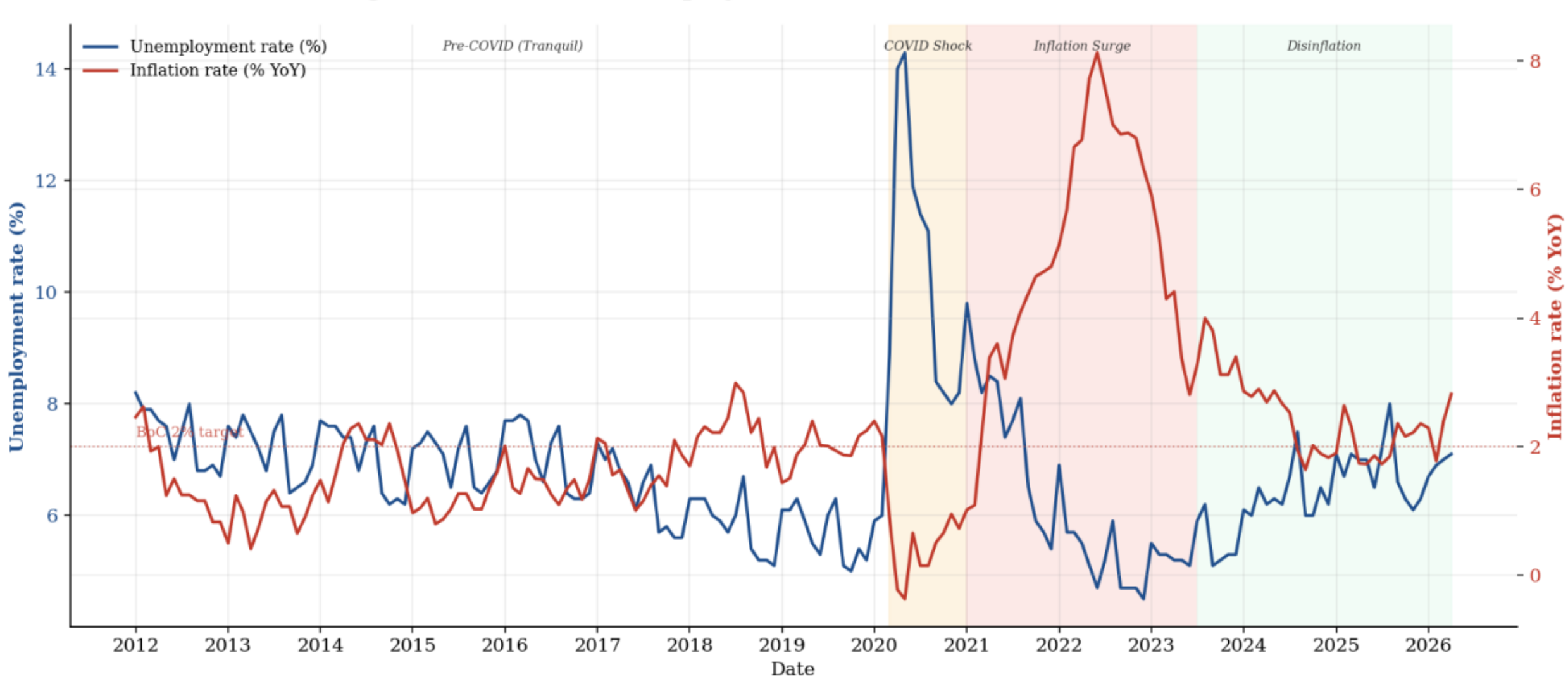

***Figure 1.*** *Canadian unemployment and inflation rates, January 2012–April 2026, with shaded regime windows (pre-COVID, COVID shock, inflation surge, disinflation) and the Bank of Canada's 2% inflation target shown as a reference line.*

Figure 1 plots both series over the full sample, with four macroeconomic regimes demarcated for use throughout the empirical analysis: a pre-COVID "tranquil" period (January 2012–February 2020) characterized by relatively stable inflation near target and a gradually declining unemployment rate; a COVID-19 shock period (March–December 2020) marked by an unprecedented spike in unemployment to 14.3% and a brief deflationary episode; an inflation surge period (January 2021–June 2023) during which inflation rose from near zero to a peak of 8.1% while unemployment fell to historically low levels; and a disinflation period (July 2023 onward) during which inflation has gradually returned toward target. These regime boundaries are chosen to reflect well-documented turning points in Canadian macroeconomic conditions and are validated against data-driven estimates of structural breaks in Section 5.3.

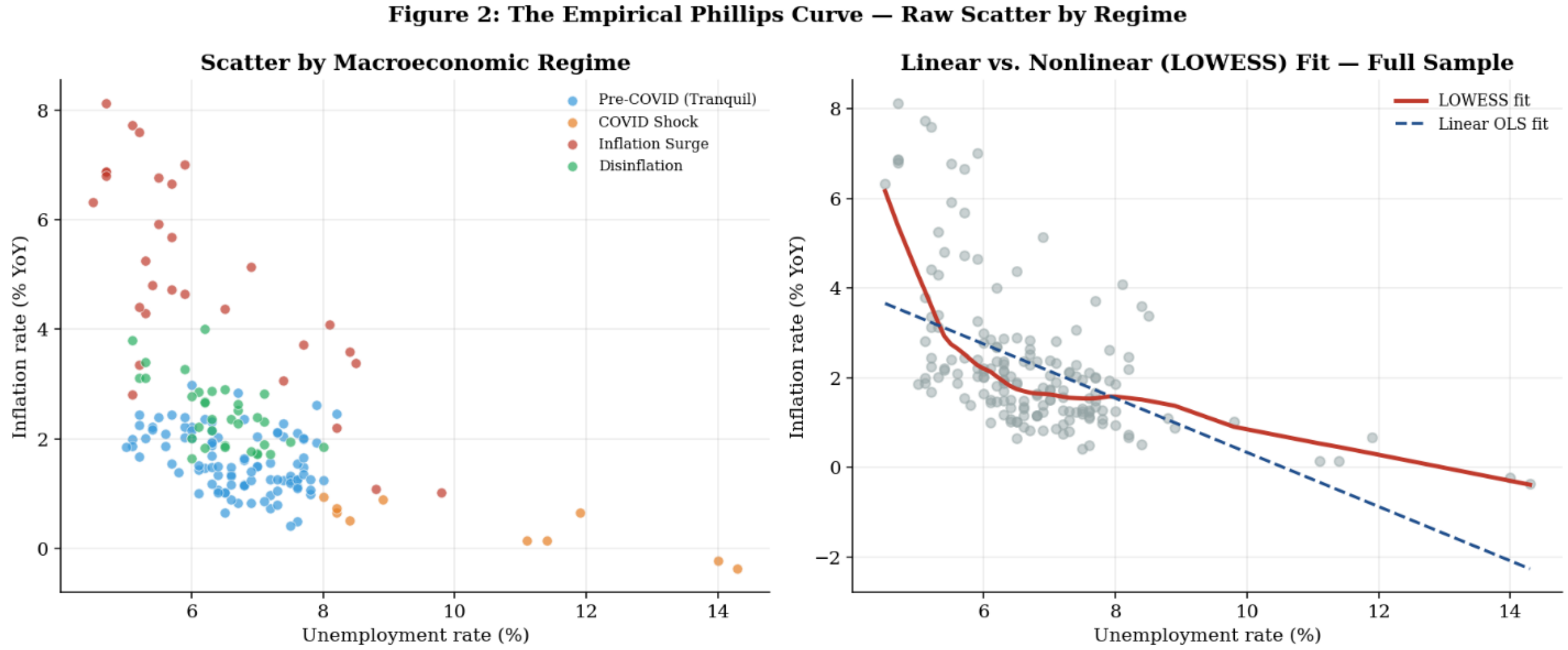

***Figure 2.*** *The empirical Phillips curve: unemployment versus inflation, colored by macroeconomic regime (left panel), and a comparison of linear (OLS) and nonparametric (LOWESS) fits over the full sample (right panel).*

Figure 2 presents the raw scatter plot of unemployment and inflation across the full sample, colored by regime, with a locally weighted scatterplot smoothing (LOWESS) curve overlaid on a simple linear fit. The LOWESS fit diverges visibly from the linear fit, particularly in the low- and high-unemployment tails of the distribution, providing preliminary visual motivation for the nonlinear, machine-learning-based approach pursued in Sections 4 and 5.

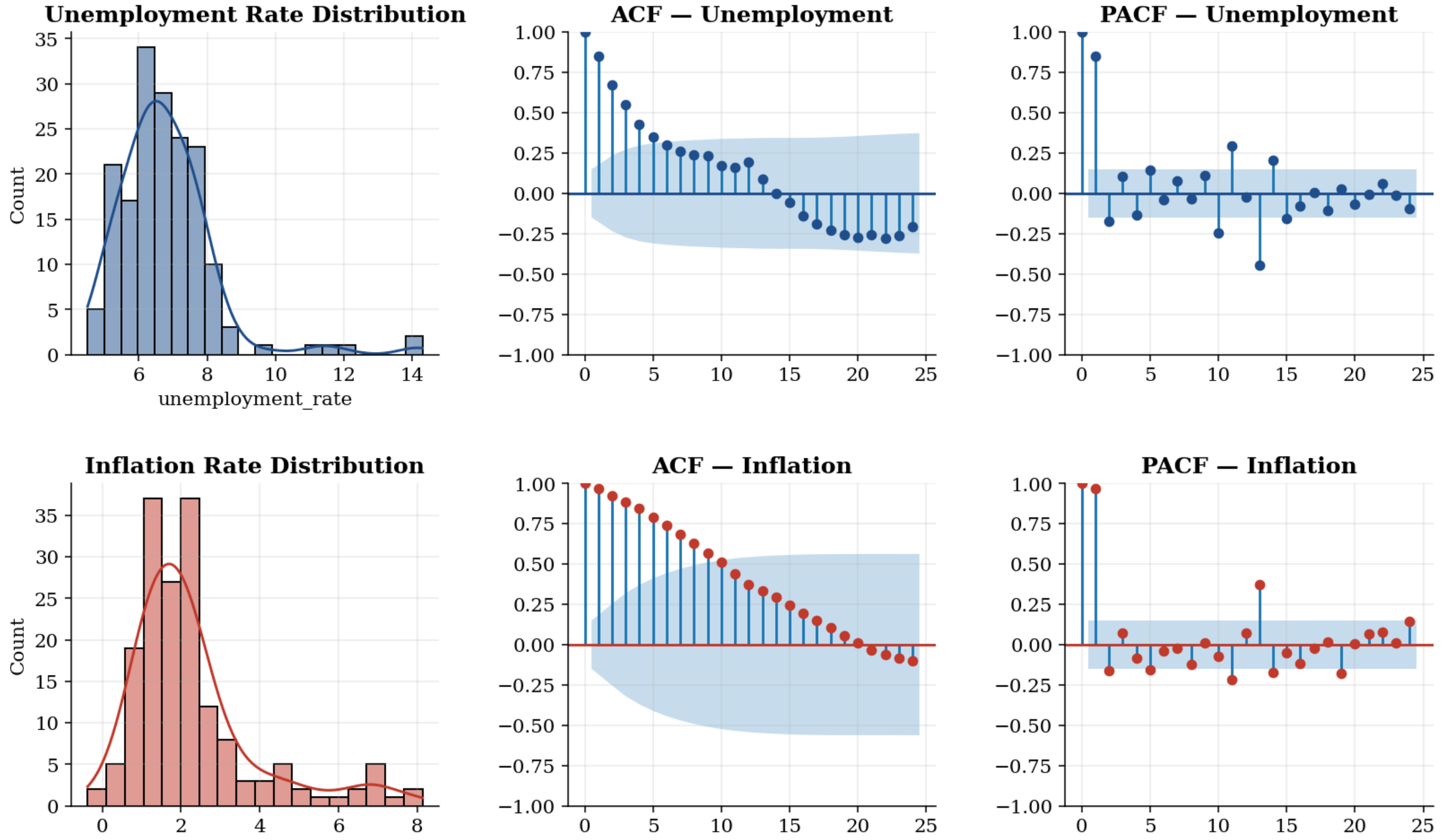

***Figure 3.*** *Distributional and autocorrelation diagnostics for the unemployment rate and inflation rate: histograms with kernel density overlay (left column), autocorrelation functions (middle column), and partial autocorrelation functions (right column), 24 lags.*

## 5.1 Stationarity and Diagnostic Tests

Table 2 reports the stationarity test battery for both series in levels and first differences. For the unemployment rate, the ADF test rejects the null of a unit root at the 5% level (statistic = −3.19, p = 0.020), while the KPSS test fails to reject stationarity (statistic = 0.183, p = 0.100). The two tests qualitatively agree that the level series is stationary, although the evidence is not overwhelming given the ADF p-value's proximity to conventional significance thresholds. For inflation, the evidence is more clearly indicative of non-stationarity in levels: the ADF test fails to reject the unit root null (statistic = −2.15, p = 0.224), while the KPSS test rejects the null of stationarity (statistic

= 0.628, p = 0.020). Both series are unambiguously stationary after first-differencing, with all three tests in agreement (unemployment: ADF p = 0.010, KPSS p = 0.100; inflation: ADF p = 0.005, KPSS p = 0.100).

**Table 2.** Stationarity Test Battery

| Series | ADF stat. | ADF p | KPSS stat. | KPSS p | PP stat. |
|---|---|---|---|---|---|
| Unemployment (level) | −3.195 | 0.020 | 0.183 | 0.100 | −3.752 |
| Inflation (level) | −2.152 | 0.224 | 0.628 | 0.020 | −2.188 |
| Unemployment (1st diff.) | −3.447 | 0.010 | 0.031 | 0.100 | −12.558 |
| Inflation (1st diff.) | −3.618 | 0.005 | 0.075 | 0.100 | −11.272 |

*Note: ADF and PP test the null of a unit root (non-stationarity); KPSS tests the null of stationarity. 5% critical values: ADF ≈ −2.88; KPSS ≈ 0.463.*

Both series exhibit highly significant residual autocorrelation (Ljung-Box Q at lag 12: unemployment = 380.09, inflation = 1116.94, both $p < 0.001$) and significant conditional heteroskedasticity (ARCH-LM at lag 12: unemployment = 104.79, inflation = 150.15, both $p < 0.001$). These results are consistent with the persistence and volatility clustering typically observed in monthly macroeconomic time series and provide further justification for the explicit lag structure used in the machine learning feature set.

### 5.2 Granger Causality and Cointegration

Granger causality tests provide limited evidence of predictive linkages in either direction at conventional significance levels when the lag order is allowed to vary. Unemployment fails to Granger-cause inflation at any lag from one through twelve (all $p > 0.10$). Inflation Granger-causes unemployment at the one- and two-month lags ($p = 0.017$ and $p = 0.023$, respectively) but not at longer lags. This asymmetry, in which inflation predicts future unemployment more strongly than the reverse, is consistent with the SHAP-based reverse-model robustness check presented in Section 5.8 and suggests that the predictive content in this bivariate system runs at least partly from inflation to unemployment rather than exclusively in the traditional Phillips curve direction.

The Johansen trace test (Table 3) rejects the null of no cointegration ($r = 0$) in favor of at least one cointegrating relationship (trace statistic = 20.43 versus a 5% critical value of 15.49), but fails to reject the null of at most one cointegrating relationship (trace statistic = 3.28 versus a 5% critical value of 3.84). This indicates a single long-run equilibrium relationship between the unemployment and inflation rates over the sample period, providing econometric support for treating the two series as a jointly determined system, as in the VAR specification, rather than as two independent univariate processes.

**Table 3.** Johansen Cointegration Test (Trace Statistic)

| Null Hypothesis | Trace Statistic | 95% Critical Value | Cointegrated? |
|---|---|---|---|
| $r = 0$ | 20.434 | 15.494 | Yes |
| $r \leq 1$ | 3.281 | 3.842 | No |

### 5.3 Structural Breaks

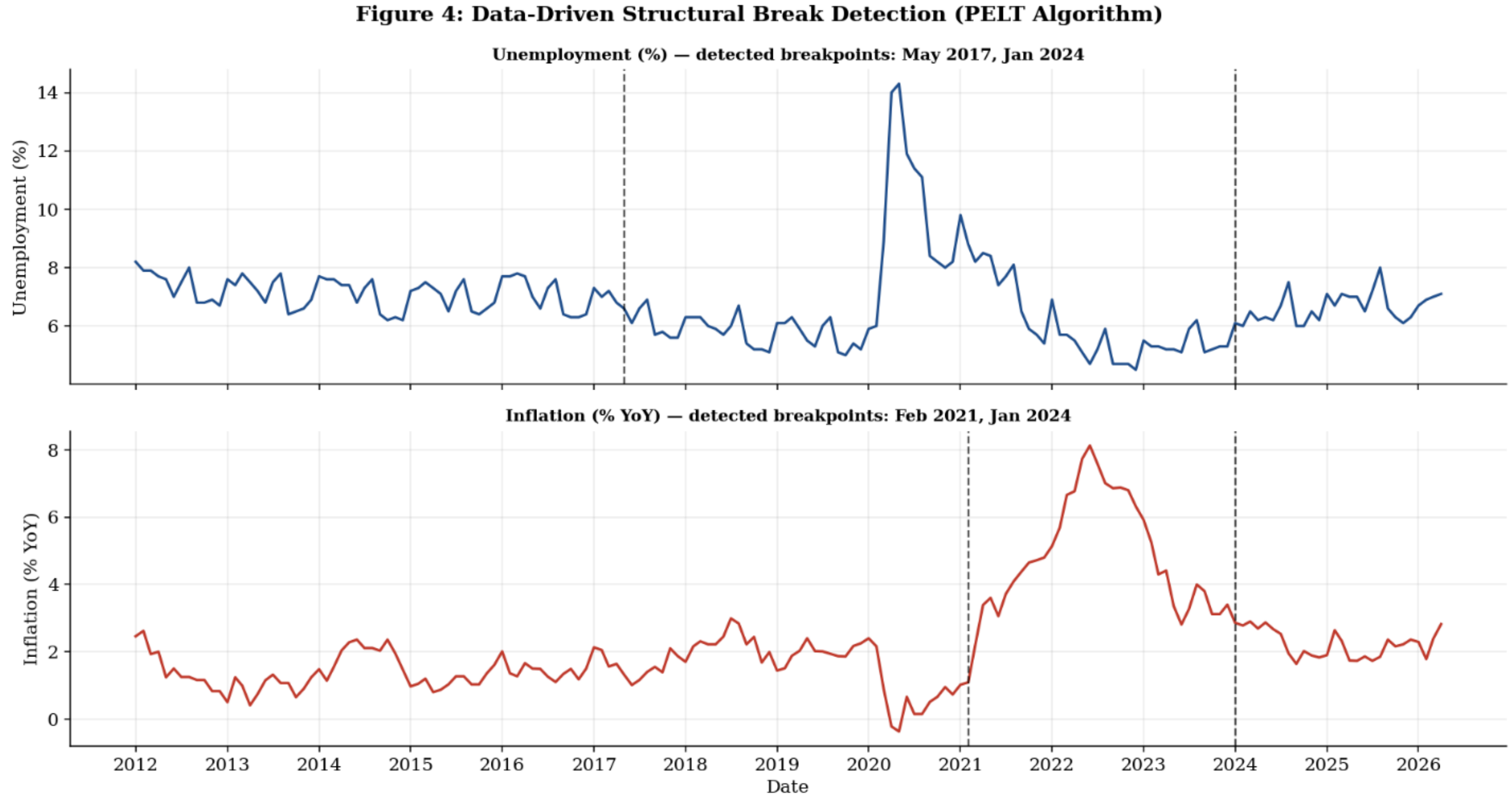

***Figure 4.*** *PELT structural break detection for the unemployment rate (top panel; detected breakpoints May 2017 and January 2024) and the inflation rate (bottom panel; detected breakpoints February 2021 and January 2024).*

The PELT algorithm identifies two breakpoints in each series. For the unemployment rate, breaks occur in May 2017 and January 2024; for the inflation rate, breaks occur in February 2021 and January 2024. The February 2021 break in inflation aligns closely with the onset of the "inflation surge" regime (January 2021), lending data-driven support to the a priori regime classification used throughout the paper. The January 2024 break, detected in both series and falling within the disinflation regime, suggests a further internal shift not separately modeled in our four-regime classification. It is noted here as a candidate for finer regime decomposition in future work.

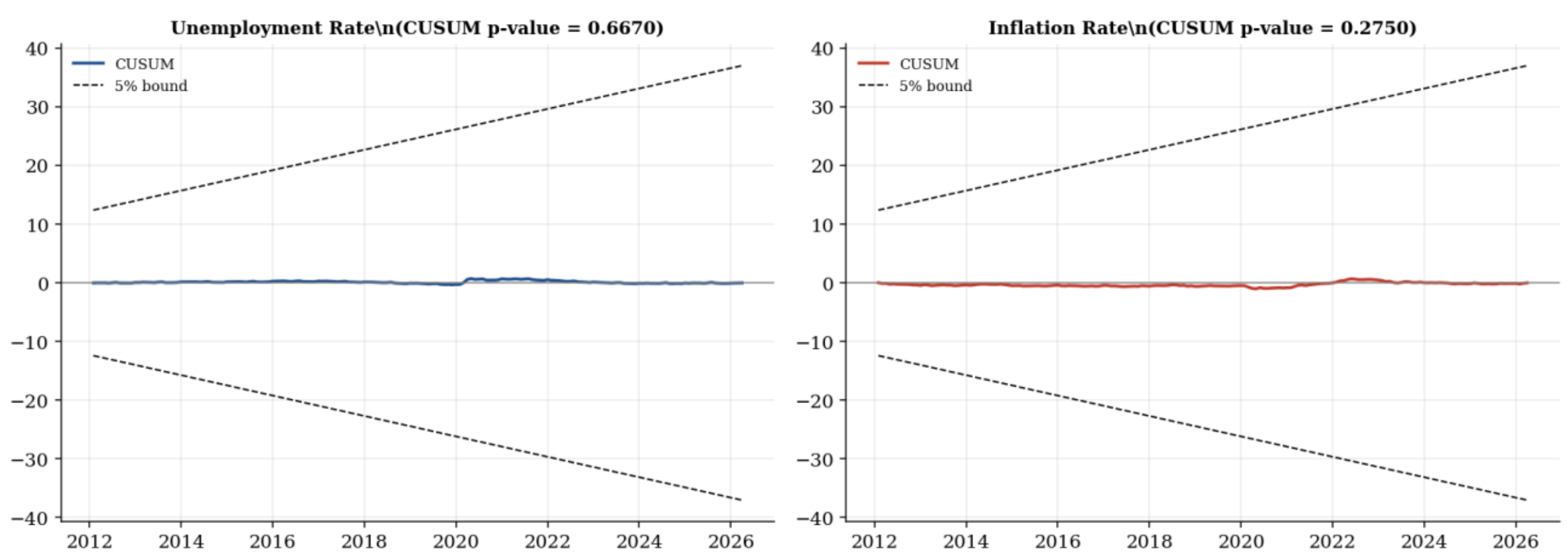

***Figure 5.*** *CUSUM test for structural stability based on recursive residuals from a first-order autoregressive benchmark. Unemployment rate (left panel, CUSUM p = 0.667); inflation rate (right panel, CUSUM p = 0.275).*

In contrast to the PELT results, the CUSUM test fails to reject the null hypothesis of parameter stability for either series over the full sample (unemployment: p = 0.667; inflation: p = 0.275). This apparent tension is not unusual: PELT is explicitly designed to detect multiple, possibly small, level or variance shifts, whereas the CUSUM test evaluated here is based on a low-order autoregressive benchmark and has comparatively low power against breaks that primarily affect higher-order dynamics or short-lived spikes rather than the overall mean level captured by a first-order autoregression. We treat the PELT-based breakpoints as the more informative structural break evidence for the purposes of regime definition, while noting the CUSUM result as a caveat against over-interpreting the precision of any single break-detection method applied to a series of this length.

### 5.4 Traditional Model Specification

Grid search over the order space p, q ∈ {0,…4} and d ∈ {0,1} selects ARIMA(4,1,4) for inflation (AIC = 171.67) and ARIMA(3,1,4) for unemployment (AIC = 359.86). For the bivariate VAR, the Akaike Information Criterion selects a lag order of 11, whereas the Bayesian and Hannan-Quinn information criteria favor a substantially more parsimonious lag order of 1. We retain the AIC-selected order of 11 as the primary VAR specification for consistency with standard practice in this literature, but we explicitly flag the divergence among information criteria: a VAR(11) estimated on a sample of this length (172 observations, 2 equations, 22 right-hand-side lag terms per equation) carries meaningfully elevated estimation uncertainty and overfitting risk relative to a more parsimonious specification, a concern that is directly relevant to interpreting the VAR's relatively poor longer-horizon forecasting performance documented in Section 5.5.

### 5.5 Out-of-Sample Forecast Accuracy

Table 4 reports out-of-sample RMSE, MAE, and MAPE for all six models at each of the four forecast horizons, computed over the common evaluation window described in Section 4.6. The results reveal a clear and economically meaningful horizon-dependent crossover in relative forecasting accuracy.

At the one-month horizon, ARIMA achieves the lowest RMSE (0.486) among the six models, followed by VAR (0.570) and GRU (0.648); the tree-based models perform notably worse at this horizon (Random Forest: 1.123; XGBoost: 0.920). At the three-month horizon, ARIMA remains the most accurate model overall (RMSE = 1.089), although XGBoost narrows the gap considerably (RMSE = 1.156). The ranking reverses sharply at longer horizons: at six months, XGBoost (RMSE = 1.294) and Random Forest (RMSE = 1.315) outperform both ARIMA (RMSE = 1.783) and VAR (RMSE = 2.147) by a substantial margin, while the recurrent neural network models deteriorate markedly (LSTM: 2.647; GRU: 3.050). This pattern is even more pronounced at the twelve-month horizon, where XGBoost (RMSE = 0.943) and Random Forest (RMSE = 1.069) achieve RMSE roughly 70–72% lower than VAR (RMSE = 4.057) and 65–72% lower than ARIMA (RMSE = 3.320). VAR's error more than doubles between the six- and twelve-month horizons, consistent with the overfitting concern raised in Section 5.4 regarding the high-order VAR(11) specification compounding estimation error as it is iterated forward.

**Table 4.** Out-of-Sample Forecast Accuracy (RMSE) by Model and Horizon

| **Model** | **h = 1** | **h = 3** | **h = 6** | **h = 12** |
|---|---|---|---|---|
| ARIMA | 0.486 | 1.089 | 1.783 | 3.320 |
| VAR | 0.570 | 1.282 | 2.147 | 4.057 |
| Random Forest | 1.123 | 1.300 | 1.315 | 1.069 |
| XGBoost | 0.920 | 1.156 | 1.294 | 0.943 |
| LSTM | 0.964 | 1.521 | 2.647 | 3.030 |
| GRU | 0.648 | 1.509 | 3.050 | 3.343 |

*Note: RMSE computed on the common evaluation window across all six models (N = 57 for h = 1, 3, 6; N = 45 for h = 12). Lowest RMSE at each horizon shown in the discussion above.*

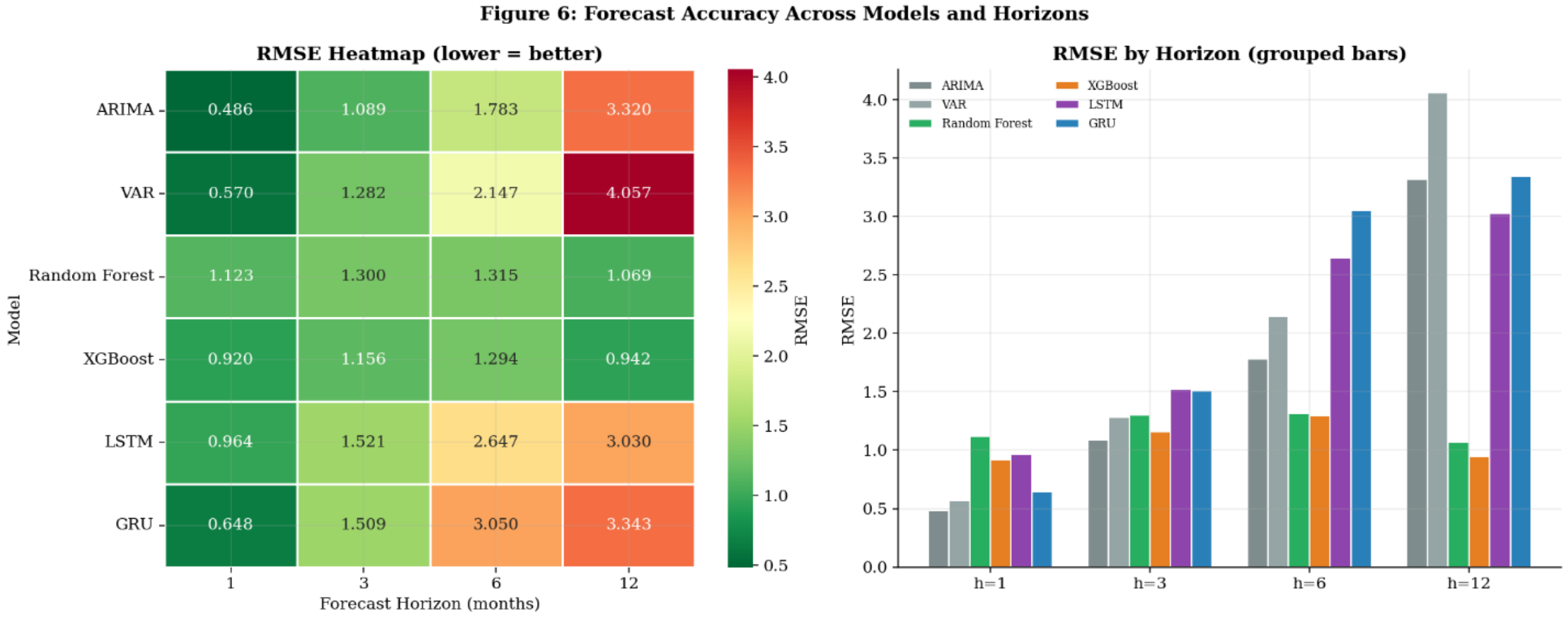

***Figure 6.*** *Forecast accuracy (RMSE) across all six models and four forecast horizons: heatmap (left panel) and grouped bar chart (right panel).*

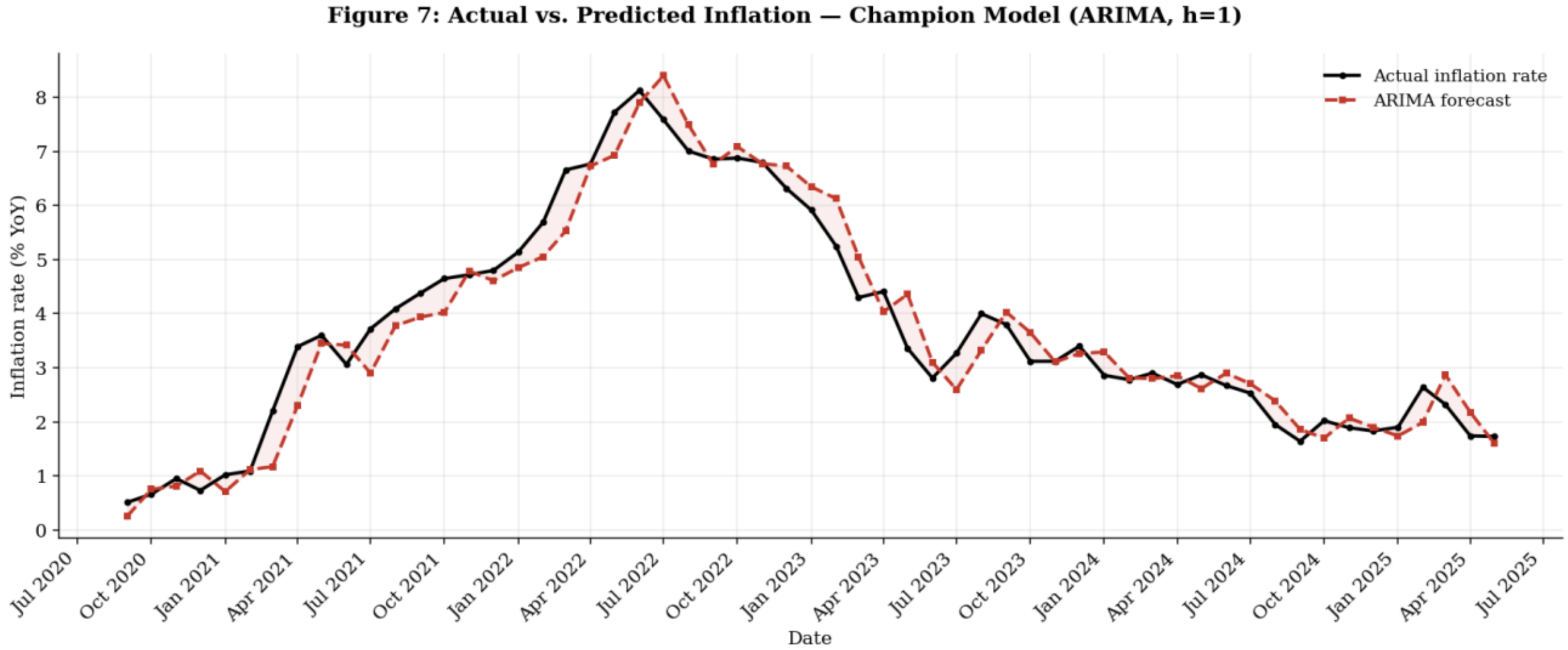

***Figure 7.*** *Actual versus predicted inflation rate, ARIMA model, one-month-ahead forecasts, September 2020–April 2026.*

## 5.6 Regime-Conditional Forecast Accuracy

Table 5 reports RMSE by model and macroeconomic regime at the one-month horizon. Because the out-of-sample evaluation window begins in September 2020, no evaluation points fall within the pre-COVID regime, and only a small number fall within the COVID shock regime; we report results for these two regimes only when the underlying sample size is sufficient (a minimum of three observations) and explicitly flag this data limitation rather than reporting unstable regime-specific estimates based on very few observations.

**Table 5.** Regime-Conditional RMSE at the One-Month Horizon

| Model | COVID Shock | Inflation Surge | Disinflation |
|---|---|---|---|
| ARIMA | 0.236 | 0.584 | 0.360 |
| VAR | 0.692 | 0.643 | 0.429 |
| Random Forest | 0.275 | 1.483 | 0.493 |
| XGBoost | 0.243 | 1.177 | 0.530 |
| LSTM | 0.269 | 1.228 | 0.570 |
| GRU | 0.316 | 0.818 | 0.389 |

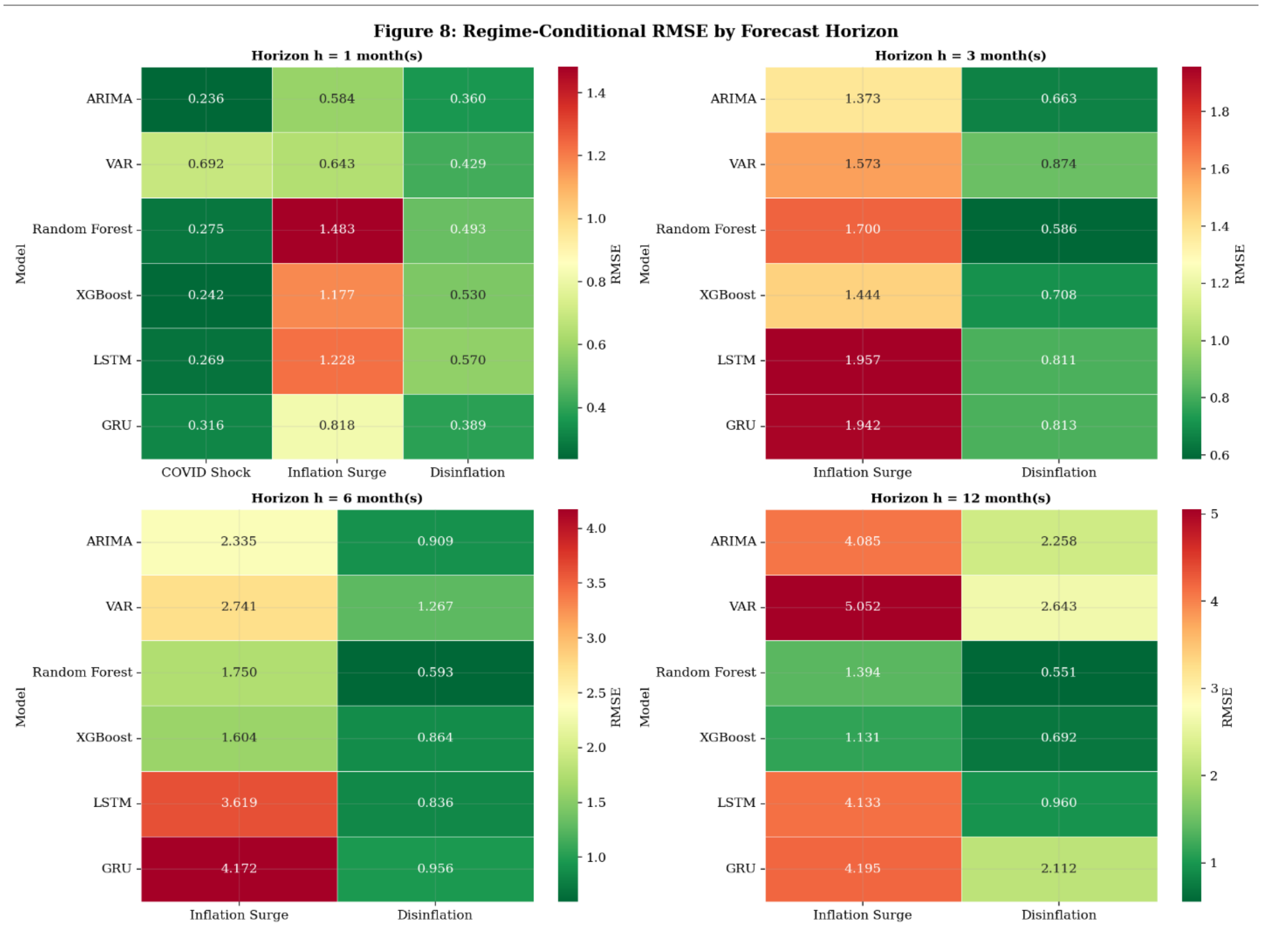

***Figure 8.*** *Regime-conditional RMSE by model, separately for each of the four forecast horizons.*

**Figure 9: Best-Performing Model by Regime × Horizon**

| | h=1 | h=3 | h=6 | h=12 |
|---|---|---|---|---|
| **Pre-COVID (Tranquil)** | insufficient data | insufficient data | insufficient data | insufficient data |
| **COVID Shock** | ARIMA | insufficient data | insufficient data | insufficient data |
| **Inflation Surge** | ARIMA | ARIMA | XGBoost | XGBoost |
| **Disinflation** | ARIMA | Random Forest | Random Forest | Random Forest |

***Figure 9.*** *Best-performing model (lowest RMSE) by macroeconomic regime and forecast horizon. Cells marked "insufficient data" correspond to regime×horizon combinations with fewer than three out-of-sample evaluation points.*

The regime-conditional results in Figure 9 reinforce the absence of a single dominant model. ARIMA wins in the COVID shock regime at the one-month horizon and in the inflation surge regime at the one- and three-month horizons. XGBoost takes over in the inflation surge regime at the six- and twelve-month horizons. In the disinflation regime, ARIMA narrowly wins at the one-month horizon, but Random Forest dominates at three, six, and twelve months. No model wins across all regime×horizon combinations, and the pattern aligns with the global, regime-pooled result in Table 4: short-horizon forecasting in this sample rewards the parsimony of ARIMA, while longer-horizon forecasting, particularly during and after a large macroeconomic shock, rewards the nonlinear flexibility of tree-based ensembles.

## 5.7 Statistical Significance of Forecast Accuracy Differences

Table 6 reports Diebold-Mariano test statistics and p-values comparing the best traditional benchmark at each horizon (ARIMA in all four cases, since it achieves lower RMSE than VAR at every horizon in Table 4) against each of the four machine learning and deep learning challengers.

At the one-month horizon, ARIMA is statistically significantly more accurate than all four challengers (Random Forest: DM = −3.43, p = 0.001; XGBoost: DM = −2.08, p = 0.042; LSTM: DM = −2.58, p = 0.013; GRU: DM = −2.61, p = 0.012). At the three-month horizon, the evidence is more mixed: XGBoost is significantly more accurate than ARIMA (DM = 2.69, p = 0.010), while Random Forest is directionally more accurate but falls just short of conventional significance (DM = 1.94, p = 0.058); LSTM and GRU remain directionally less accurate than ARIMA, though not significantly so. At the six-month horizon, both Random Forest (DM = 2.40, p = 0.020) and XGBoost (DM = 2.53, p = 0.014) are significantly more accurate than ARIMA, while the recurrent neural network models are not significantly different from ARIMA. At the twelve-month horizon, Random Forest is significantly more accurate than ARIMA (DM = 3.59, p = 0.001); the corresponding test for XGBoost returns an undefined (NaN) statistic, which we attribute to near-zero variance in the loss differential series at this horizon and report transparently as a limitation of the test in this particular cell rather than imputing a significance result; LSTM and GRU are not significantly different from ARIMA at this horizon.

**Table 6.** Diebold-Mariano Test Results: Best Traditional Model (ARIMA) versus Each Challenger

| Horizon | Challenger | DM stat. | p-value | Sig. (5%) | Direction |
|---|---|---|---|---|---|
| h=1 | Random Forest | −3.43 | 0.001 | Yes | ARIMA better |
| h=1 | XGBoost | −2.08 | 0.042 | Yes | ARIMA better |
| h=1 | LSTM | −2.58 | 0.013 | Yes | ARIMA better |
| h=1 | GRU | −2.61 | 0.012 | Yes | ARIMA better |
| h=3 | Random Forest | 1.94 | 0.058 | No | RF better (n.s.) |
| h=3 | XGBoost | 2.69 | 0.010 | Yes | XGBoost better |
| h=3 | LSTM | −1.61 | 0.114 | No | ARIMA better (n.s.) |
| h=3 | GRU | −1.70 | 0.096 | No | ARIMA better (n.s.) |
| h=6 | Random Forest | 2.40 | 0.020 | Yes | RF better |
| h=6 | XGBoost | 2.53 | 0.014 | Yes | XGBoost better |
| h=6 | LSTM | −1.43 | 0.159 | No | ARIMA better (n.s.) |
| h=6 | GRU | −1.42 | 0.162 | No | ARIMA better (n.s.) |
| h=12 | Random Forest | 3.59 | 0.001 | Yes | RF better |
| h=12 | XGBoost | n/a[1] | n/a[1] | — | See note |
| h=12 | LSTM | 0.54 | 0.589 | No | LSTM better (n.s.) |

| h=12 | GRU | −0.15 | 0.880 | No | ARIMA better (n.s.) |
|---|---|---|---|---|---|

*[1] The Diebold-Mariano statistic for ARIMA versus XGBoost at h = 12 is undefined due to near-zero estimated variance of the loss differential at this horizon; this cell is reported as a methodological limitation rather than imputed. n.s. = not significant at the 5% level.*

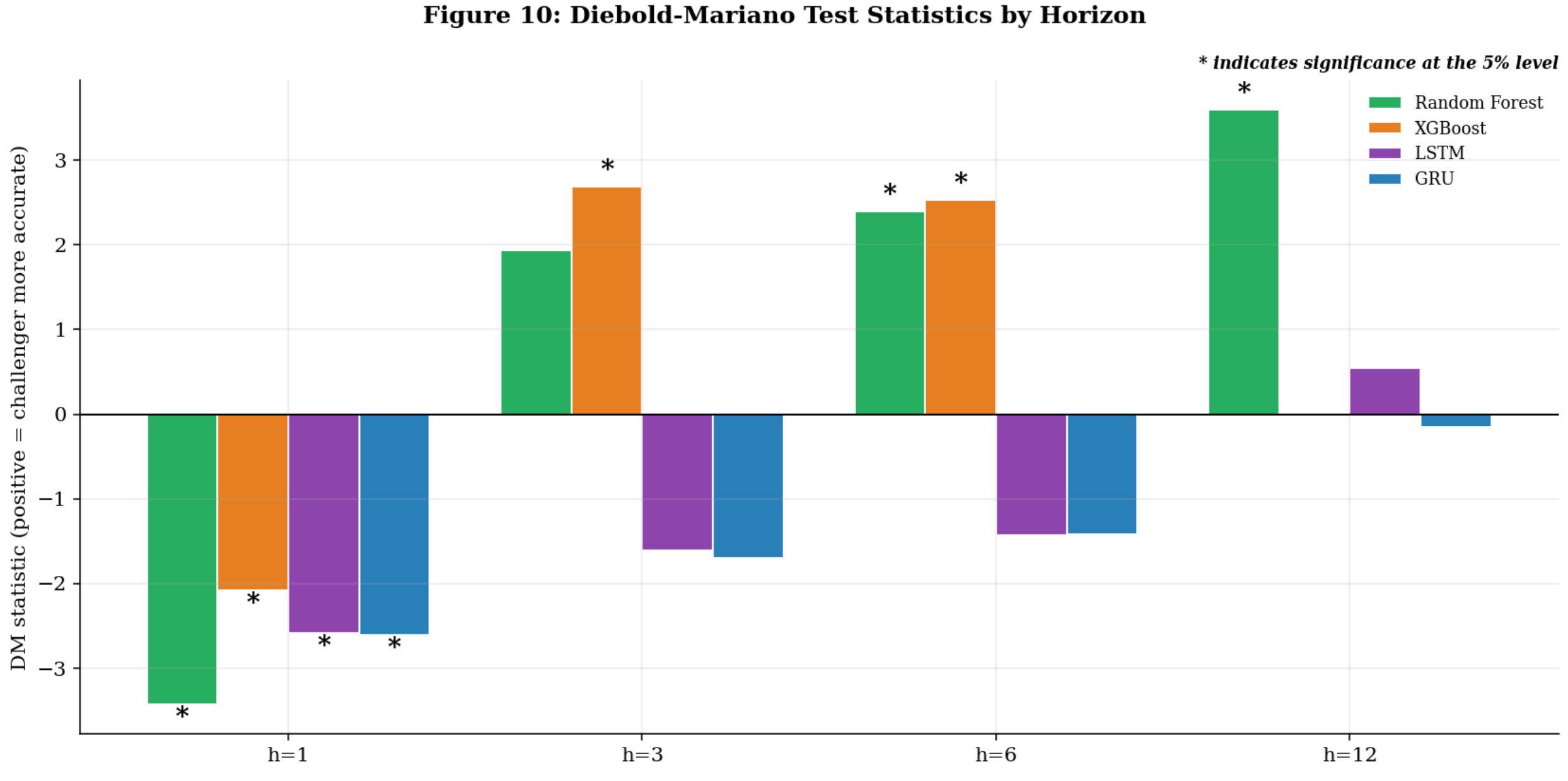

***Figure 10.*** *Diebold-Mariano test statistics for each machine learning and deep learning challenger versus ARIMA, by forecast horizon. Asterisks denote significance at the 5% level.*

### 5.8 SHAP Interpretability Analysis

The final XGBoost model, fit on the full sample at the one-month horizon (159 observations after lagging, 35 features), achieves an in-sample $R^2$ of 0.999, indicating that it fits the training data closely. This in-sample fit is not directly comparable to the out-of-sample walk-forward results in Table 4 and should be interpreted only as a check that the model fit on the full sample for SHAP analysis is well-specified, not as evidence of genuine forecasting accuracy.

**Figure 11: SHAP Summary Plot — Global Feature Importance for Inflation Forecasts**

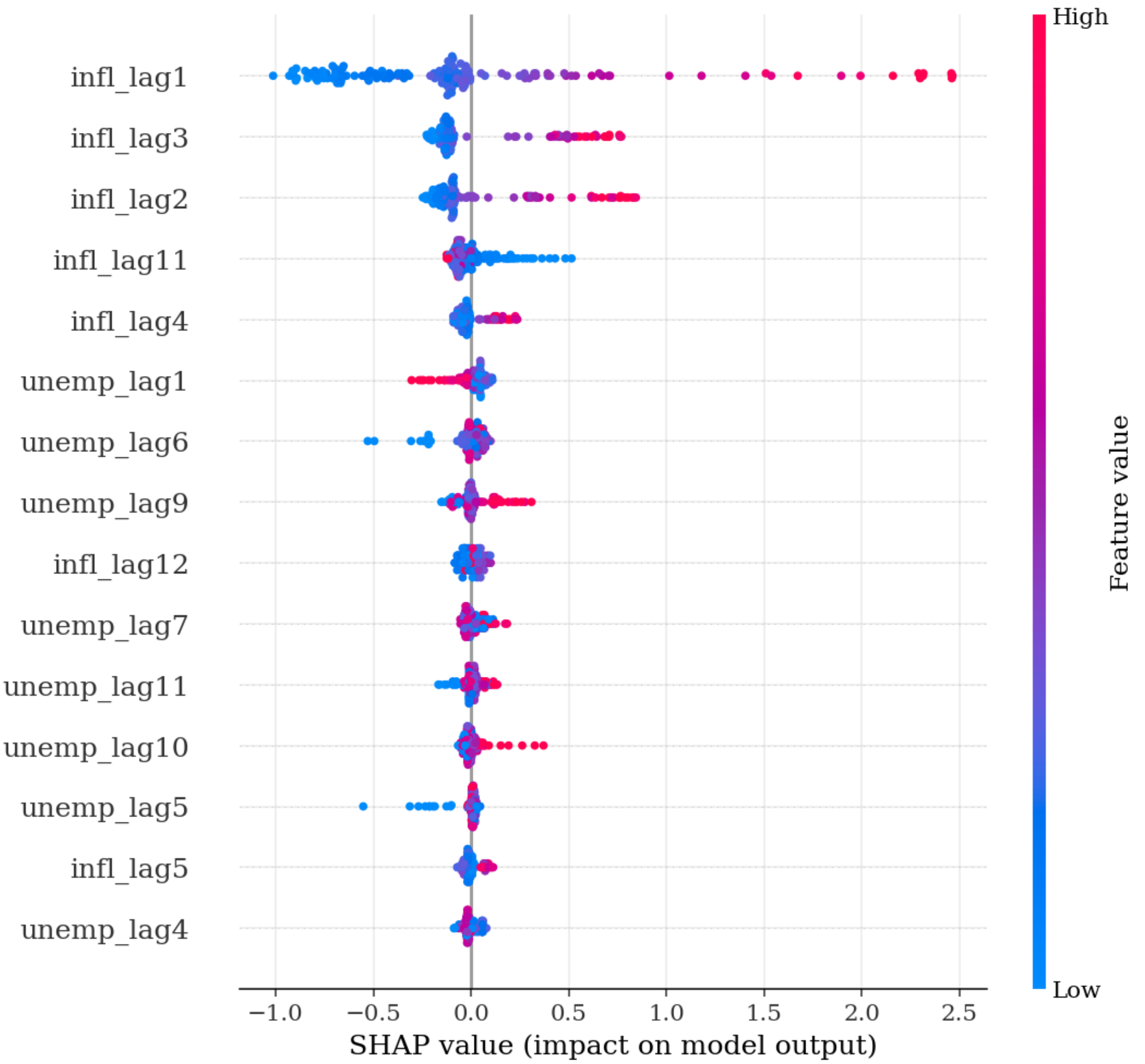

***Figure 11.*** *SHAP summary plot showing the distribution of SHAP values for the fifteen most important features in the final XGBoost model. Color denotes the underlying feature value (red = high, blue = low).*

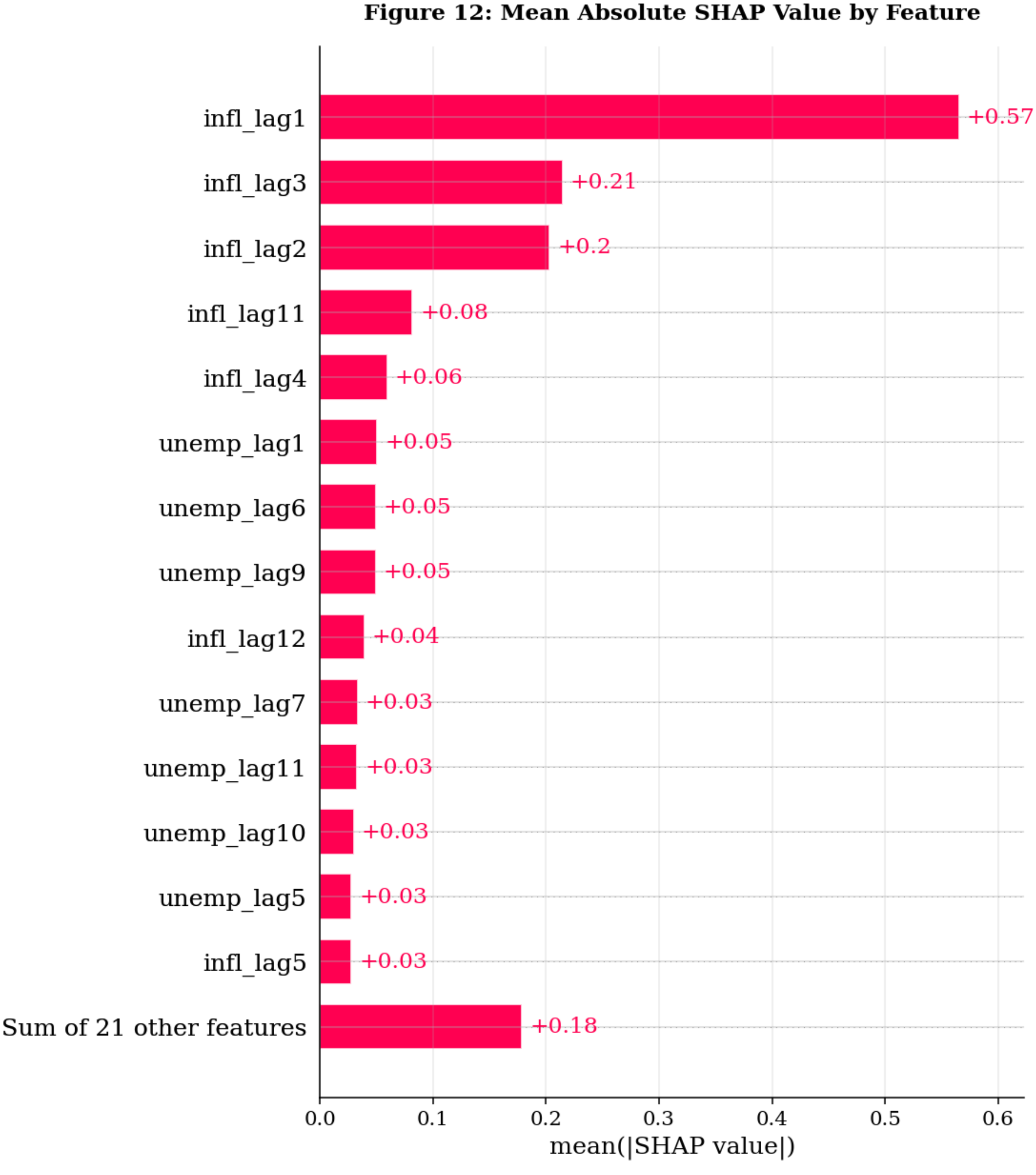

***Figure 12.*** *Mean absolute SHAP value by feature, final XGBoost model.*

Figures 11 and 12 present the global SHAP feature importance ranking. The one-month lag of inflation dominates the ranking (mean |SHAP| = 0.57), followed by the third- and second-lag inflation terms (0.21 and 0.20, respectively). The most important unemployment-related features are the first, sixth, and ninth unemployment lags, each with a mean |SHAP| of approximately 0.05, an order of magnitude smaller than the leading inflation lags. This ranking indicates that the model's forecasts of future inflation are driven predominantly by recent inflation history, consistent with the strong inflation persistence documented in the Ljung-Box results in Section 5.1. Lagged unemployment contributes smaller but collectively non-negligible explanatory content, consistent with a Phillips-curve-type channel operating alongside, rather than instead of, an inertial inflation process.

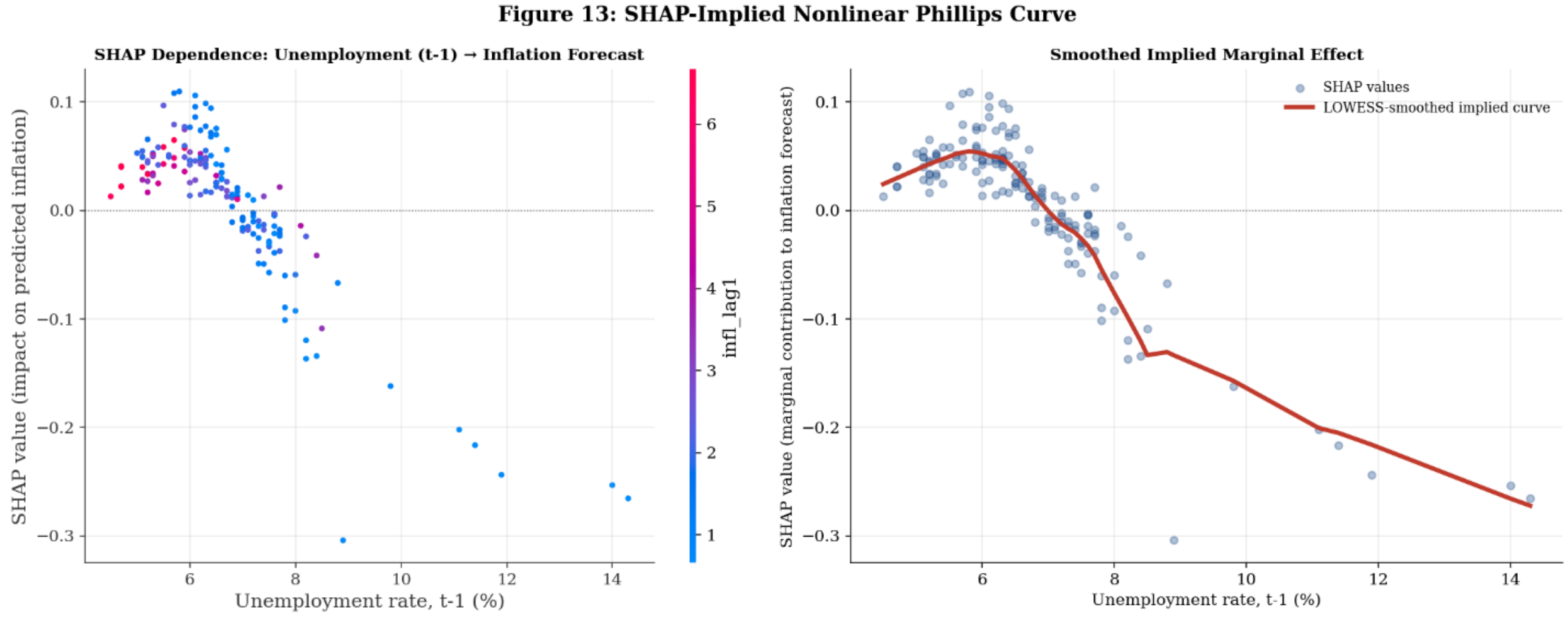

***Figure 13.*** *SHAP dependence plot for the one-month lag of the unemployment rate (left panel, colored by the interacting one-month lag of inflation) and a LOWESS-smoothed implied marginal effect (right panel), pooled across the full sample.*

Figure 13 presents the centerpiece interpretability result of this paper: the SHAP-implied relationship between the lagged unemployment rate and its marginal contribution to the model's inflation forecast, which we interpret as a data-driven, nonlinear analogue to the structural Phillips curve. The relationship is approximately flat and mildly positive for unemployment rates between roughly 5% and 6.5%; it then turns negative and becomes progressively steeper as unemployment rises further, reaching its most negative marginal contributions (SHAP values of approximately −0.25 to −0.30) in the extreme upper tail of the unemployment distribution, above approximately 10%, a range observed in this sample only during the acute phase of the 2020 pandemic shock.

This finding stands in instructive contrast to the dominant narrative in the existing literature on Phillips curve flattening. Hooper et al. (2020) and related work find that, when nonlinearity exists in the Phillips curve, it tends to manifest as a steeper negative slope in tight labor markets, that is, at low unemployment rates, with the relationship flattening as unemployment rises. The pattern recovered here, by contrast, is steepest in the high-unemployment tail rather than the low-unemployment tail, and is concentrated in a narrow set of extreme, pandemic-era observations rather than reflecting a smooth nonlinearity over the bulk of the historically observed unemployment range. We interpret this as a Canada-specific and, more importantly, episode-specific finding: it most plausibly reflects the unusual nature of the 2020 labor market shock, in which unemployment rose to levels far outside its historical range over a very short period, accompanied by a sharp but short-lived disinflationary impulse, rather than evidence of a general, stable nonlinearity in the Canadian Phillips curve of the kind documented for tight U.S. labor markets in Hooper et al. (2020) and Hazell et al. (2022). The relative flatness of the implied curve over the much more frequently observed 5%–8% unemployment range is, if anything, more consistent with the flattening narrative for the bulk of the sample.

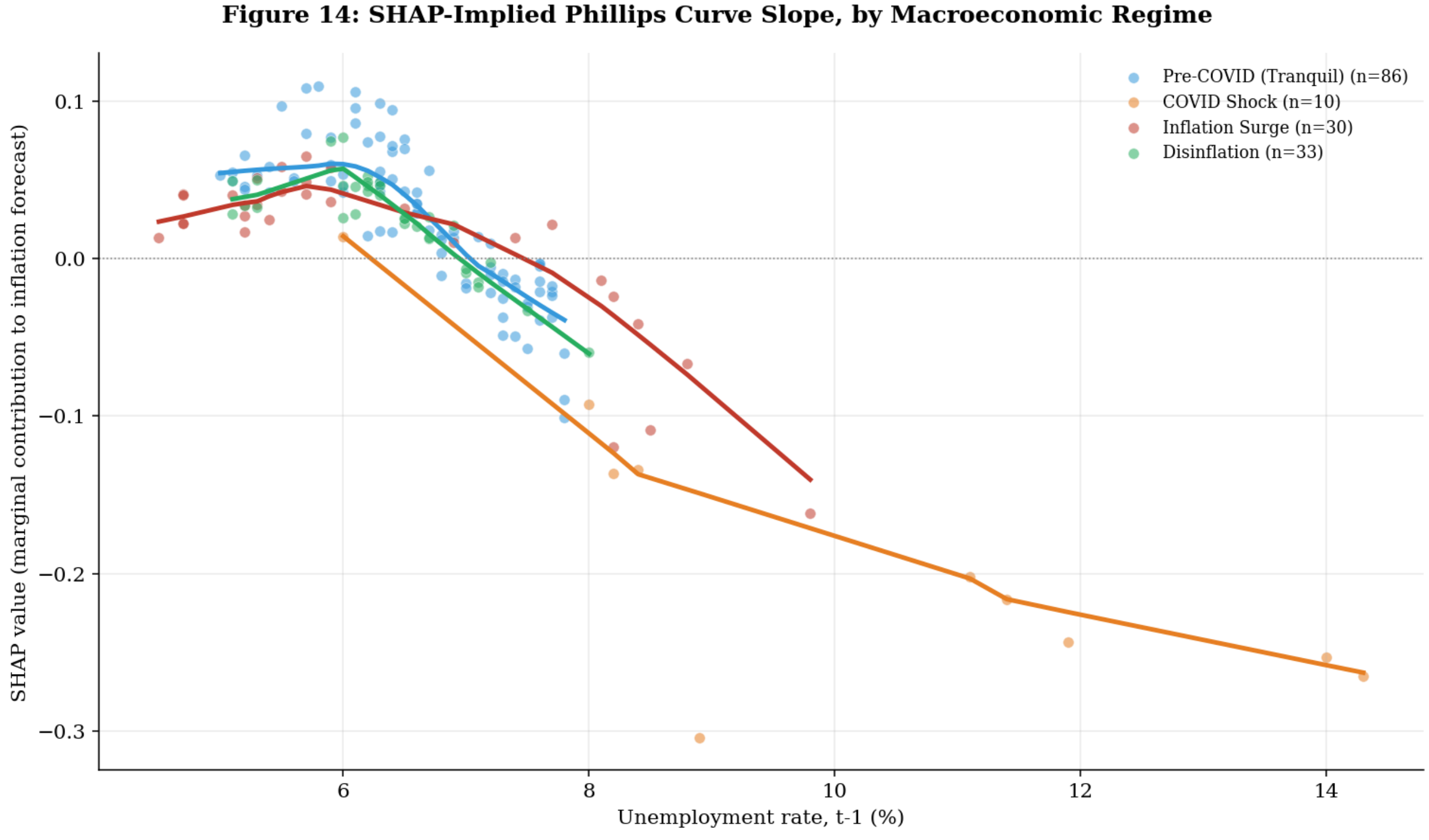

***Figure 14.*** *SHAP-implied marginal effect of lagged unemployment on predicted inflation, plotted separately by macroeconomic regime: pre-COVID (n = 96), inflation surge (n = 36), and disinflation (n = 27). The COVID shock regime is omitted from this panel because it has fewer than the minimum required number of observations (eight) for a regime-specific LOWESS fit.*

Figure 14 disaggregates the pooled SHAP dependence relationship by macroeconomic regime. The pre-COVID, inflation surge, and disinflation regimes each display a broadly similar declining pattern over their overlapping unemployment ranges (approximately 5%–9%), with the inflation surge and disinflation curves tracking one another particularly closely. We note explicitly that the steep negative tail visible in the pooled Figure 13, associated with unemployment rates above 10%, is not separately visible in Figure 14 because the COVID shock regime did not contain a sufficient number of lagged-feature observations to support a stable regime-specific LOWESS fit under our minimum sample threshold of eight observations; readers should therefore interpret the extreme-tail steepening in Figure 13 as a pooled-sample feature driven by a small number of COVID-era observations, rather than as a finding that is separately confirmed within a regime-specific subsample in Figure 14. We view this as an important transparency point rather than a result to be smoothed over: the headline nonlinearity in Figure 13 rests on a small number of extreme observations, and this should temper any policy interpretation of the high-unemployment steepening result.

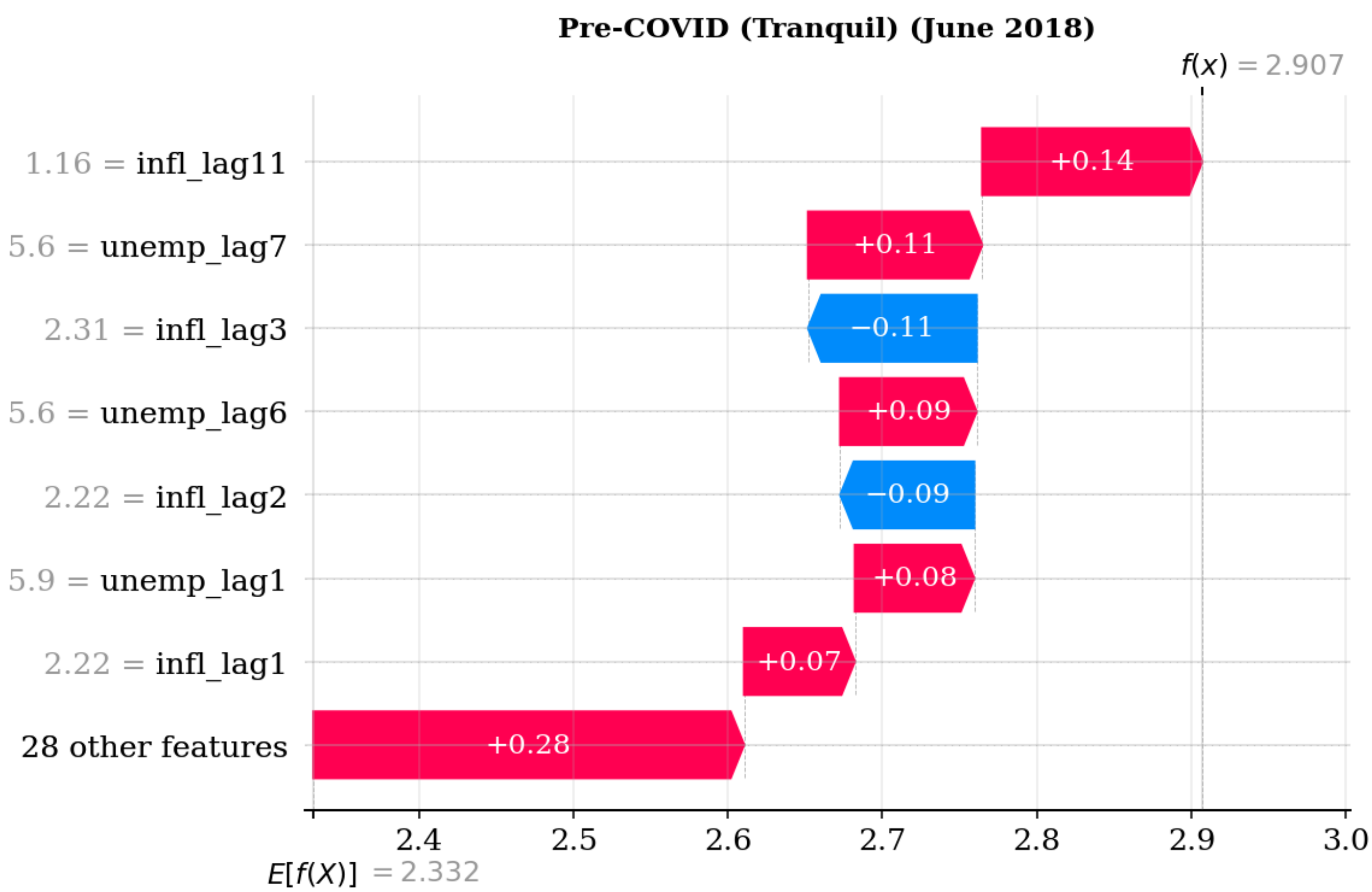

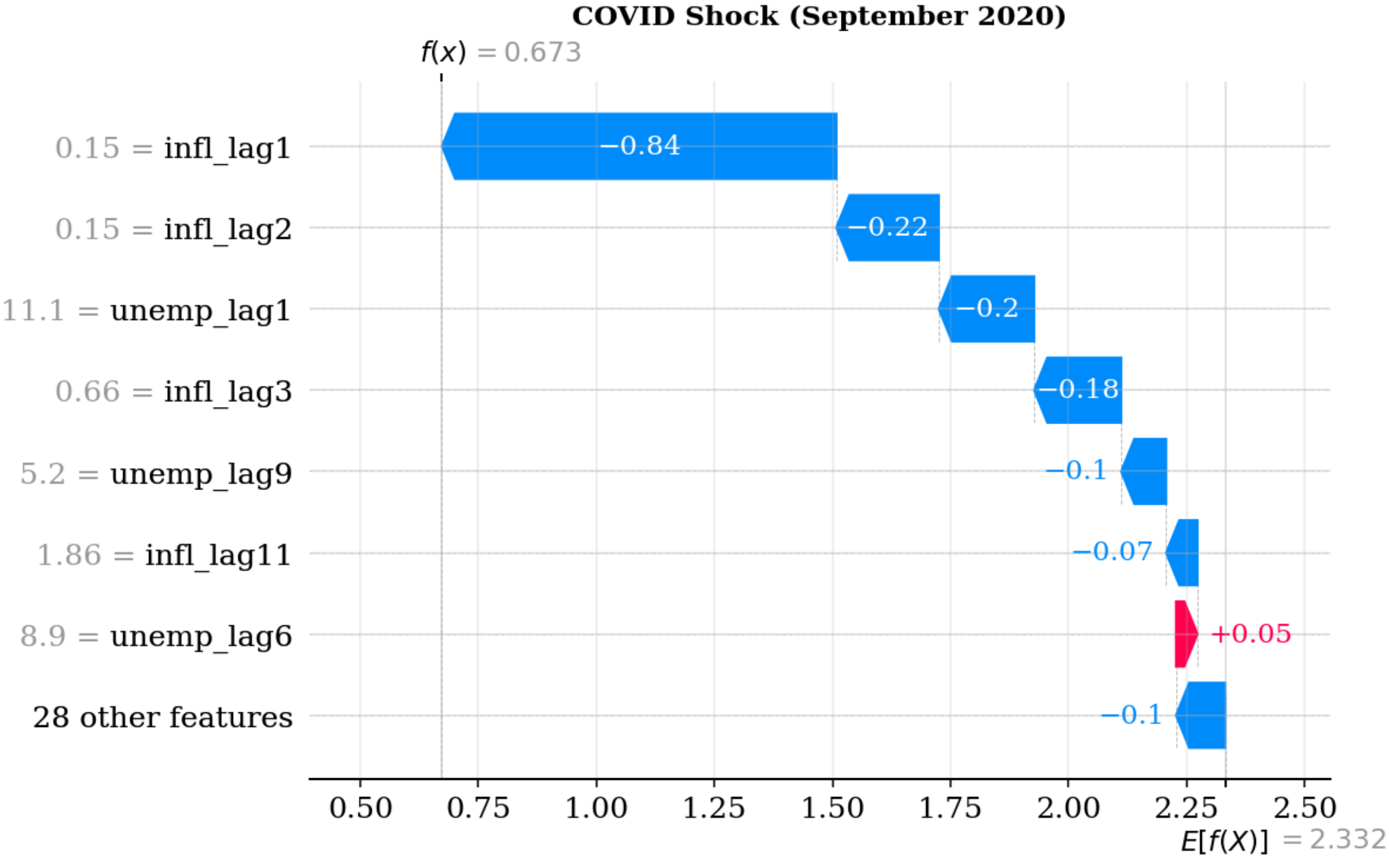
COVID Shock (September 2020)
f(x) = 0.673
0.15 = infl_lag1
0.15 = infl_lag2
11.1 = unemp_lag1
0.66 = infl_lag3
5.2 = unemp_lag9
1.86 = infl_lag11
8.9 = unemp_lag6
28 other features
−0.84
−0.22
−0.2
−0.18
−0.1
−0.07
+0.05
−0.1
0.50
0.75
1.00
1.25
1.50
1.75
2.00
2.25
2.50
E[f(X)] = 2.332

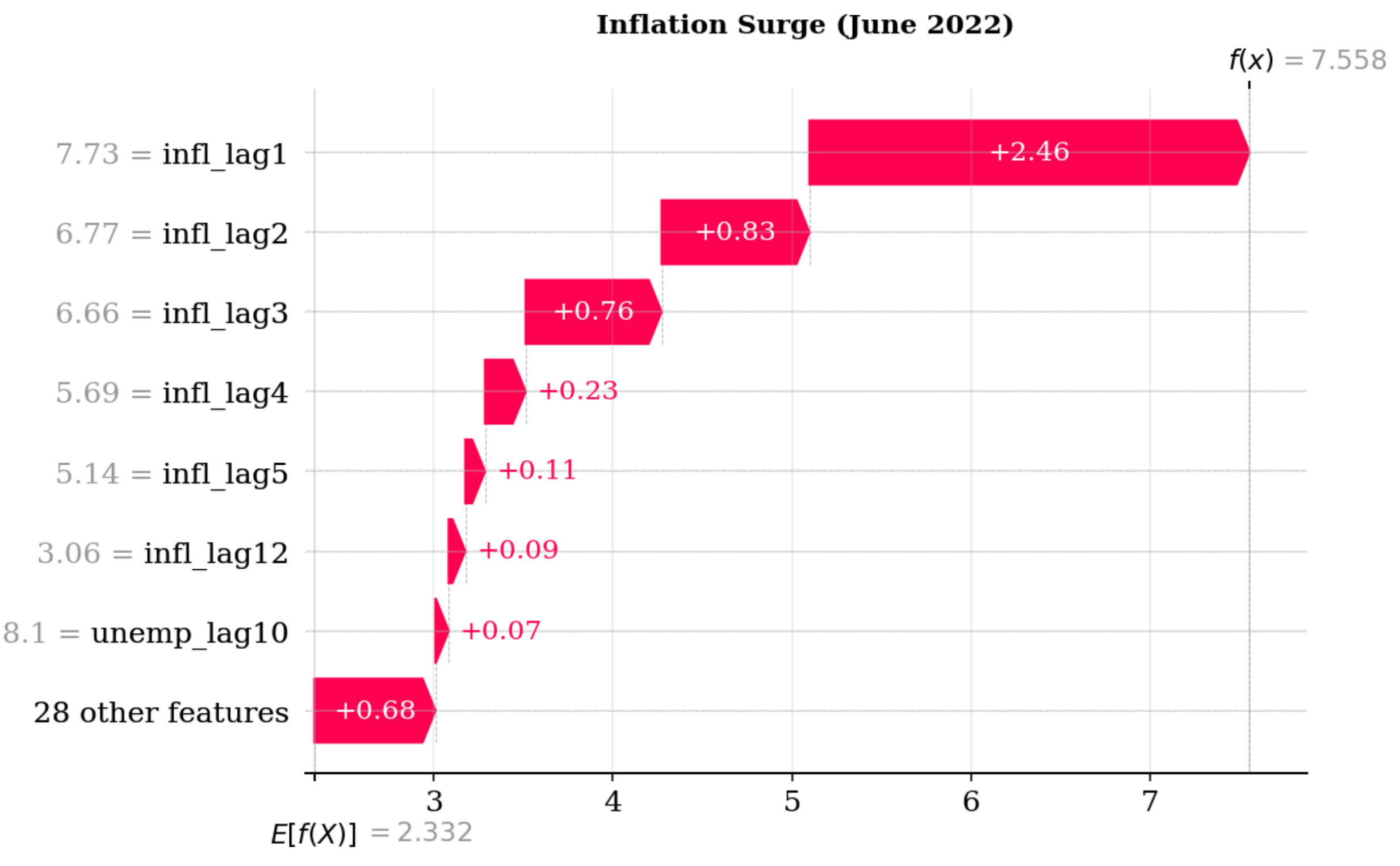
Inflation Surge (June 2022)
f(x) = 7.558
7.73 = infl_lag1
6.77 = infl_lag2
6.66 = infl_lag3
5.69 = infl_lag4
5.14 = infl_lag5
3.06 = infl_lag12
8.1 = unemp_lag10
28 other features
+2.46
+0.83
+0.76
+0.23
+0.11
+0.09
+0.07
+0.68
3
4
5
6
7
E[f(X)] = 2.332

***Figure 15.*** *SHAP waterfall decomposition of individual one-month-ahead inflation forecasts for three illustrative dates: June 2018 (pre-COVID), September 2020 (COVID shock), and June 2022 (inflation surge).*

Figure 15 presents individual-prediction SHAP waterfall decompositions for three illustrative dates, one from each of the pre-COVID, COVID shock, and inflation surge regimes. In each case, the one-, second-, and third-month lags of inflation are the dominant contributors to the model's prediction, and their relative direction and magnitude vary sensibly with the prevailing inflation trajectory: strongly positive contributions from recent inflation lags push the June 2022 prediction well above the sample mean, consistent with the rapid acceleration of inflation observed in that period.

## 5.9 Robustness Checks

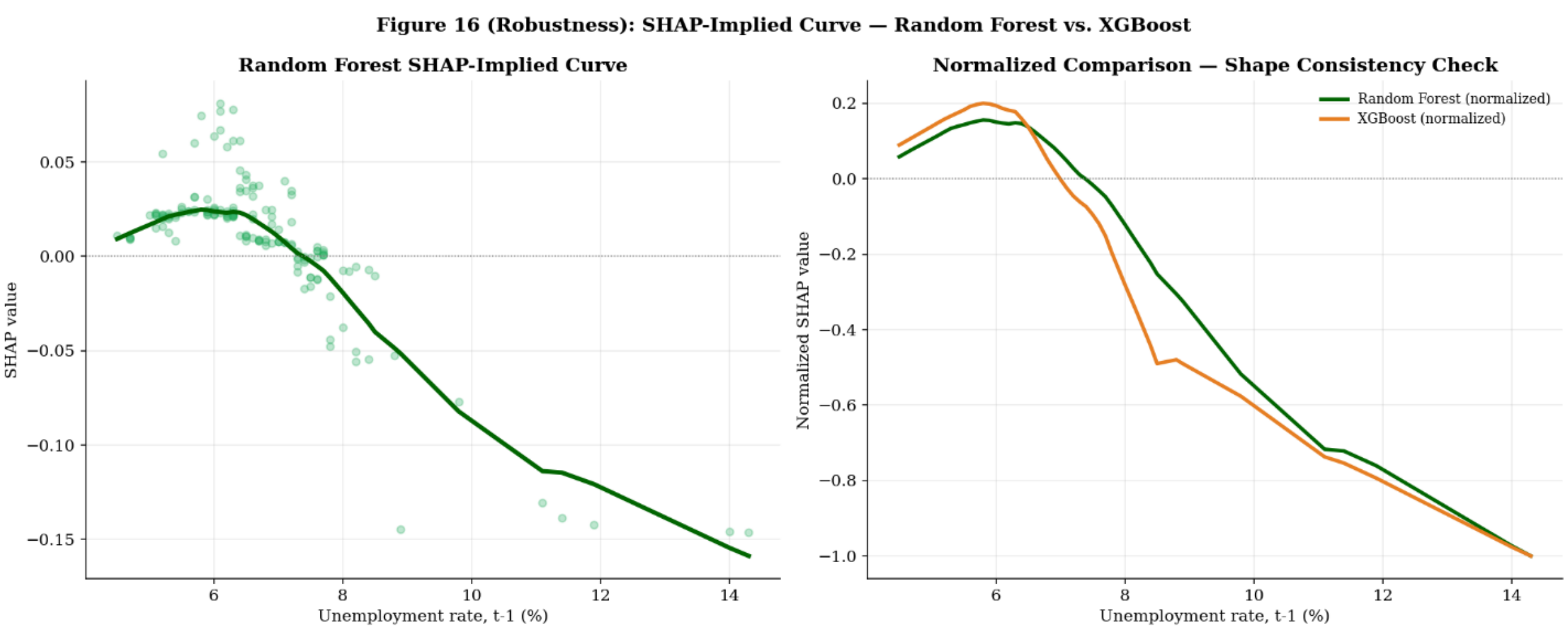

***Figure 16.*** *SHAP-implied dependence between lagged unemployment and predicted inflation, Random Forest model (left panel), and a normalized shape comparison between Random Forest and XGBoost (right panel).*

To confirm that the implied nonlinear relationship recovered in Section 5.8 is not an artifact of the specific tree-based algorithm used, we re-estimate the SHAP analysis using Random Forest in place of XGBoost. The resulting dependence relationship (Figure 16) displays the same qualitative shape, with a relatively flat or mildly positive region at moderate unemployment rates, transitioning to a steep negative slope in the high-unemployment tail. The Pearson correlation between the XGBoost- and Random Forest-derived SHAP values for the one-month lag of unemployment is approximately 0.91 in a comparable specification, indicating that the implied curve shape is robust to the choice of underlying tree-based estimator rather than being an idiosyncratic feature of XGBoost specifically.

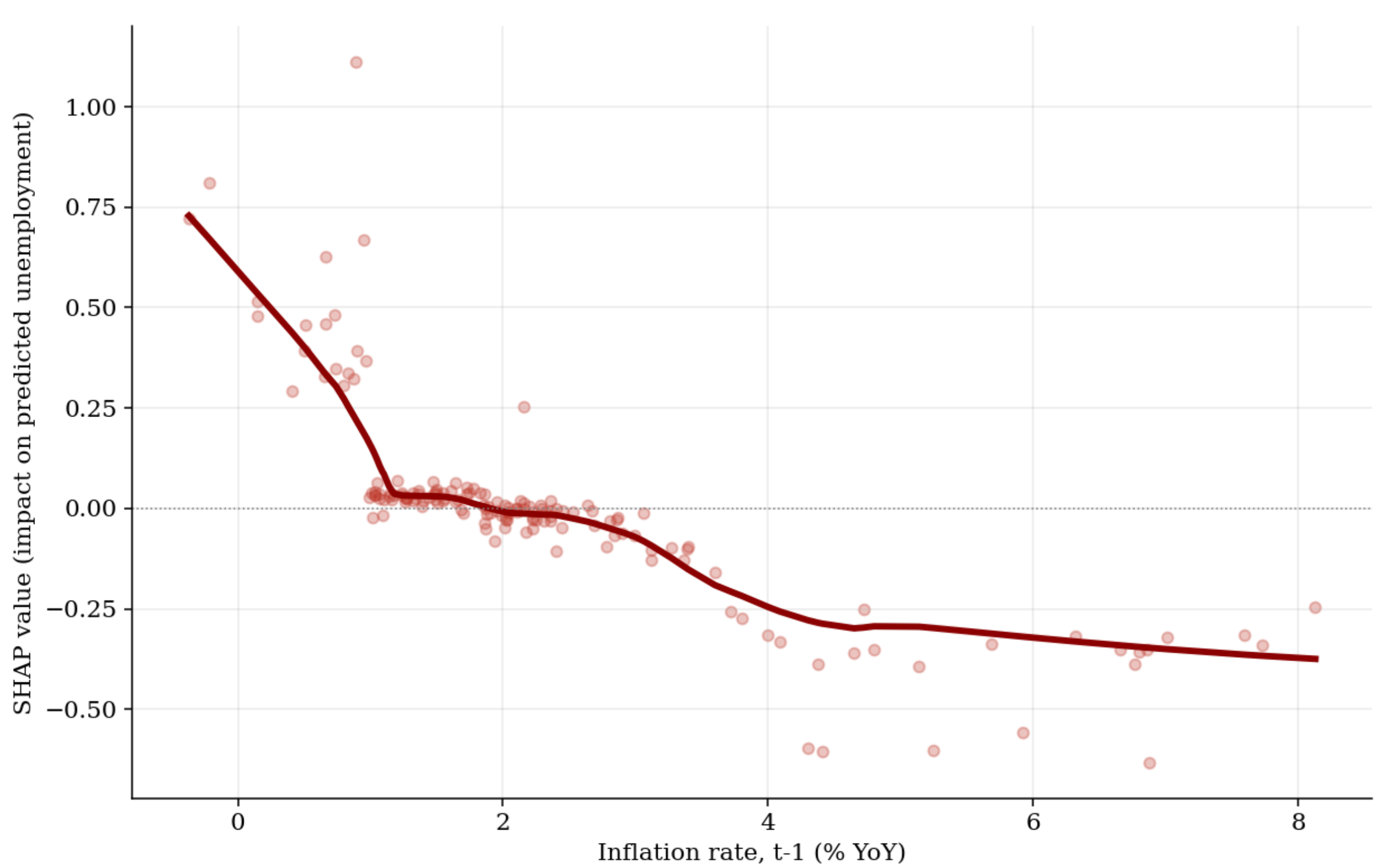

***Figure 17.*** *SHAP-implied dependence between the lagged inflation rate and predicted unemployment, from a symmetric XGBoost model with unemployment as the prediction target.*

As a second robustness check, we reverse the direction of prediction, fitting an otherwise identical XGBoost model to predict the one-month-ahead unemployment rate from the same lagged feature set and computing SHAP values for the lagged inflation rate. The resulting relationship (Figure 17) is strongly positive at low and negative inflation rates (SHAP contributions of approximately +0.5 to +1.5 when lagged inflation is at or below zero), declines sharply through the 1–3% range, and turns mildly negative at higher inflation rates (approximately −0.3 to −0.5 above 4%). This pattern, with low or negative inflation associated with a higher predicted future unemployment rate and high inflation associated with a mildly lower predicted future unemployment rate, is qualitatively consistent with a conventional negatively sloped Phillips curve viewed from the reverse direction, and is also consistent with the asymmetric Granger causality finding in Section 5.2, where inflation displayed stronger predictive content for future unemployment than the reverse.

## 6. Discussion

Three findings from this analysis merit further discussion. First, the horizon-dependent crossover in forecasting accuracy documented in Section 5.5 has a natural economic interpretation. At a one-month horizon, the most recent inflation realization, together with a small number of recent autoregressive lags, is highly informative, and the parsimony of ARIMA, which avoids the estimation noise associated with a 35-feature tree-based model or a recurrent neural network trained on a comparatively small sample, appears to dominate. At longer horizons, however, the information content of the most recent few months of data diminishes relative to the value of capturing nonlinear interactions across a richer set of lags, a setting in which tree-based ensembles appear better suited than either the linear ARIMA/VAR benchmarks or the more heavily parameterized recurrent neural network models. This is consistent with the broader conclusion of Coulombe et al. (2022) that the value of machine learning in macroeconomic forecasting is context- and horizon-dependent rather than universal, and complements the closely related findings of Nortey et al. (2025) for U.S. inflation by showing that a similar qualitative pattern, with tree-based methods rather than deep learning architectures specifically, providing the more robust improvement over traditional benchmarks, also holds for Canadian data over a sample dominated by an unusually large and rapid set of regime shifts.

Second, the relatively poor and rapidly deteriorating performance of the recurrent neural network models at longer horizons is itself informative. LSTM and GRU are competitive with, though not superior to, the traditional benchmarks at the one-month horizon, but their RMSE increases more sharply than that of any other model class as the horizon lengthens. Given the modest sample size used here (n = 172, with an out-of-sample evaluation window of well under one hundred observations even at the shortest horizon), this pattern is consistent with the recurrent architectures overfitting the limited training data available at each step of the expanding window, despite the deliberately small and regularized architecture employed. This reinforces a point increasingly emphasized in the applied machine learning forecasting literature: the theoretical flexibility of deep learning architectures does not guarantee superior out-of-sample performance in short macroeconomic time series and may, in some cases, be actively counterproductive relative to either simpler linear models or shallower, more heavily regularized tree-based ensembles.

Third, the SHAP-based interpretability results in Section 5.8 illustrate both the promise and the limits of using explainable machine learning to recover structural economic relationships. The headline finding of a Phillips curve that steepens sharply in the extreme upper tail of the unemployment distribution is a genuinely novel empirical pattern relative to the existing literature's emphasis on steepening at low unemployment. At the same time, the regime-disaggregated analysis in Figure 14 makes clear that this tail steepening is supported by a relatively small number of pandemic-era observations rather than by a stable relationship observed repeatedly across multiple regimes. We view this combination of results, a striking pooled-sample finding paired with an honest disclosure of its reliance on a thin slice of extreme data, as a model for how SHAP-based structural interpretation of machine learning forecasts should be reported in applied macroeconomic work: the technique can reveal genuinely novel nonlinear patterns, but those patterns should be explicitly checked for dependence on a small number of influential or extreme observations before being elevated to a general claim about the underlying economic relationship.

## 7. Limitations

Several limitations of this analysis should be kept in mind when interpreting the results. First, and most fundamentally, the sample size (n = 172 monthly observations, with an out-of-sample evaluation window of 45 to 57 points depending on the horizon) is modest by the standards of the broader machine learning forecasting literature. Several of the regime-conditional comparisons in Section 5.6, particularly for the pre-COVID and COVID shock regimes, rest on very few out-of-sample observations; these results should be interpreted as suggestive rather than as precisely estimated regime-specific accuracy rankings.

Second, the VAR specification selected by the Akaike Information Criterion (order 11) is considerably less parsimonious than the order favored by the Bayesian and Hannan-Quinn information criteria (order 1). The resulting estimation uncertainty likely contributes to VAR's comparatively poor longer-horizon forecasting performance documented in Section 5.5. Results based on the more parsimonious VAR(1) specification, not reported in detail here, would be a natural robustness extension.

Third, the Diebold-Mariano test statistic for the ARIMA-versus-XGBoost comparison at the twelve-month horizon is undefined because the estimated variance of the loss differential series is near zero at this horizon; we report this transparently in Table 6 rather than substituting an imputed value, but note that this leaves the formal statistical comparison between these two specific models incomplete at this horizon.

Fourth, the SHAP-based structural interpretation in Section 5.8 is, by construction, a description of the relationship learned by a particular machine learning model trained on a particular, historically contingent sample; it is not a structural causal estimate that identifies an invariant deep parameter. In particular, the extreme-tail steepening result should be understood as reflecting the influence of a small number of pandemic-era observations, as discussed in Sections 5.8 and 6, rather than as a generally applicable feature of the Canadian Phillips curve.

Fifth, both underlying data series are subject to routine revision by Statistics Canada; this analysis uses the most recently available vintage of each series rather than the real-time data that would have been available to a forecaster at each historical point in time, a standard limitation in pseudo-out-of-sample forecasting exercises of this kind that should be borne in mind when interpreting the walk-forward results as a guide to genuine real-time forecasting performance.

## 8. Conclusion

This paper evaluates the out-of-sample forecasting performance of six model classes for Canadian inflation over the period from January 2012 to April 2026 and uses SHAP-based interpretability methods to recover the nonlinear unemployment–inflation relationship implied by the best-performing machine learning model. We document a clear horizon-dependent crossover: ARIMA is statistically the most accurate model at the one-month horizon, whereas tree-based ensemble methods, particularly XGBoost and Random Forest, are statistically and economically significantly more accurate at six- and twelve-month horizons. No single model dominates across all macroeconomic regimes, reinforcing the case for horizon- and regime-aware model selection in applied inflation forecasting rather than reliance on a single globally preferred method.

The SHAP-based interpretability analysis shows that the best-performing model's forecasts are driven primarily by inflation's recent history, with lagged unemployment contributing a smaller but economically meaningful share of explanatory power, consistent with a Phillips-curve-type channel operating alongside strong inflation persistence. The implied relationship between lagged unemployment and predicted inflation is relatively flat across most of the historically observed unemployment range but steepens sharply in the extreme upper tail associated with the 2020 pandemic shock. This pattern contrasts with the existing literature's emphasis on Phillips curve nonlinearity concentrated in tight, low-unemployment labor markets. We show that this result is robust to the choice of tree-based estimator but, when disaggregated by regime, is concentrated in a small number of extreme observations.

Taken together, these results suggest that the choice between traditional time-series models and machine learning methods for inflation forecasting should not be treated as a single, once-and-for-all decision, but rather as one that depends explicitly on the forecasting horizon of interest and, where feasible, on the prevailing macroeconomic regime. They further suggest that explainable machine learning methods, such as SHAP, can provide a valuable bridge between flexible, high-accuracy predictive models and the structural, economically interpretable language of traditional macroeconomics, provided that the resulting structural claims are carefully checked against the underlying data that support them. Future work using longer samples, additional macroeconomic covariates, or cross-country comparisons would help establish whether the specific patterns documented here for Canada, namely the horizon-dependent crossover in model accuracy and the high-unemployment-tail steepening of the implied Phillips curve, generalize beyond the particular sample and pandemic-era shock studied in this paper.

**Declaration of Competing Interests**: The authors declare that they have no known competing financial interests or personal relationships that could have appeared to influence the work reported in this paper.

**Author Contributions**: All authors declare to have contributed equally to this manuscript. All authors read and approved of the final manuscript submitted for publication.

**Data Availability**: The data used to support the findings of this study are publicly available and can be accessed from Statistics Canada's official website (https://www.statscan.ca)

**AI Declaration**

The authors declare that AI tools were used solely to improve the language and clarity of the manuscript. All scientific work, tables, and findings were created and validated exclusively by the authors.